\documentclass[a4paper, 11pt]{article}
\usepackage{amsmath,amssymb,revsymb}
\usepackage[dvipdfmx]{graphicx}
\usepackage{here}
\usepackage{bm}
\usepackage{fancybox}
\usepackage{secdot}
\usepackage{comment}
\usepackage{braket}
\usepackage{titlesec}
\usepackage{geometry}
\usepackage{hyperref}
\usepackage{bookmark}
\usepackage{slashed}
\usepackage{fancyhdr}
\usepackage{boxedminipage}
\usepackage{color}
\usepackage{feynmf}
\usepackage{multirow}
\usepackage{dcolumn}
\usepackage{feynmf}%%%{feynmp}
\usepackage{float}
\usepackage[hang,small,bf]{caption}
\usepackage[subrefformat=parens,skip=20pt]{subcaption}
\renewcommand{\arraystretch}{1.0}
\makeatletter

\@addtoreset{equation}{section}
\makeatother

\makeatletter

\@addtoreset{table}{section}
\makeatother

\titleformat*{\section}{\LARGE\bfseries}
\titleformat*{\subsection}{\Large\bfseries}

\hypersetup{
setpagesize=false,
bookmarksnumbered=true,%
bookmarksopen=true,%
colorlinks=true,%
linkcolor=blue,
citecolor=blue,
}

\begin{document}

%--------------E-mail--------------%

\fancypagestyle{foot}
{
\fancyfoot[L]{$^{*}$E-mail address : hinata0903@akane.waseda.jp }
\fancyfoot[C]{}
\fancyfoot[R]{}
\renewcommand{\headrulewidth}{0pt}
\renewcommand{\footrulewidth}{0.5pt}
}

\renewcommand{\footnoterule}{%
\kern -3pt
\hrule width \columnwidth
\kern 2.6pt}

%%%%%%%%%%%%%%%%%%% TITLE PAGE %%%%%%%%%%%%%%%%%%%%%%%%%%%%%

\begin{titlepage}
\begin{flushright}
\begin{minipage}{0.2\linewidth}
\normalsize

%--------------arXiv number--------------%
WU-HEP-21-06\\
\end{minipage}
\end{flushright}

\begin{center}

\vspace*{5truemm}
\Large
\bigskip\bigskip

%title%
\LARGE\textbf{Seesaw mechanism in the R-parity violating supersymmetric standard model with the gauged flavor $\mathrm{U}(1)_X$ symmetry}%

\Large

%authority%
\bigskip\bigskip
Atsushi Hinata$^{*}$%
\vspace{1cm}

{\large \it{Department of Physics, Waseda University, Tokyo 169-8555, Japan}}\\
\bigskip\bigskip\bigskip

\large\textbf{Abstract}\\
\end{center}
%abstract%
\ \ We study the seesaw mechanism in the supersymmetric standard model (SSM) with ${\mathbb Z}_3$ symmetry instead of the R-parity, so-called {\it Matter triality} ($M_3$). This Abelian discrete symmetry prohibits the baryon/lepton number violation operators at the non-renormalizable level, and the proton longevity is ensured. The lepton number violation term by the right-handed neutrino is only allowed under the symmetry, it plays a role of the Majorana mass after the right-handed sneutrino develops into the vacuum expectation value. The mass of the active neutrino is generated from the two contributions from the right-handed neutrino and the neutralino due to the R-parity violation. In this paper, we realize the neutrino mass and mixing angle in the SSM with matter triality which is embedded into the gauged flavor ${\mathrm{U}}(1)_X$ symmetry. In addition to the flavor ansatz, we derive the charge assignment to satisfy the anomaly cancellation conditions. Then, the model could predict the sterile neutrino with the mass below the soft SUSY breaking mass scale, and its mass and mixing angle are restricted by the search for the heavy neutral leptons.

%Updated to the revised version with added the phenomenological discussion and figures.
\thispagestyle{foot}

\end{titlepage}

\baselineskip 7.5mm

%%---------------------- CONTENTS PAGE ------------------------------ %%

\tableofcontents
\clearpage

%%---------------------- MAIN ------------------------------ %%

\parindent=10pt

\setlength{\baselineskip}{18pt}

%--------section1-------------%
\section{Introduction}
The standard model (SM) in particle physics successfully explains the physics at the electroweak (EW) scale, while there remain some open problems that cannot be explained in the framework of the SM. Its supersymmetric extension has been explored as the solution to the naturalness problem that is based on the quantum correction to the mass of the SM Higgs fields because of the huge gap between the EW scale and the Planck scale \cite{Martin:1997ns}. Although the superpartner of the SM particle has not been observed in any experiment, the supersymmetric standard model (SSM) could be motivated as the low energy effective theory in the context of the superstring theory. 
  
 Even if some problems in SM can be solved by supersymmetric extension, the introduction of the supersymmetry leads to the fast proton decay via propagation of the sparticles \cite{Goto:1998qg, Murayama:1994tc, Harnik:2004yp}. This process arises from the baryon number ($B$) or lepton number ($L$) violation operators which is invariant under the SM gauge symmetry. In the minimal supersymmetric standard model (MSSM), the R-parity is assigned to prohibit these operators at the renormalizable level. 
While such $B/L$ violation operators appear again at the renormalisable level even if the R-parity is assigned, the other classes of the Abelian discrete symmetry can forbid these dangerous operators instead of the R-parity. In Refs.\cite{Ibanez:1991hv, Ibanez:1991pr}, they derive some classes of the Abelian discrete symmetries satisfying the discrete gauge anomaly cancellation condition following that the quantum gravity effects break the global symmetry \cite{Krauss:1988zc, Kallosh:1995hi}. Based on the anomaly cancellation, the supersymmetric SM (SSM) in the R-parity violation (RpV) scenario has been studied as an alternative to the MSSM \cite{Dreiner:2005rd, Lee:2007qx, Byakti:2017igi}.
 In general, the supersymmetric SM (SSM) violating R-parity allows the characteristic operators to violate $B$ or $L$. Then, the R-parity violation (RpV) scenario is applied to the phenomenology in the supersymmetry with or without further extensions \cite{Barbier:2004ez, Allanach:2003eb, Allanach:2007qc, Colucci:2018yaq}. 

The flavor problem in the Yukawa coupling is also the main issue of physics beyond the standard model (BSM). The quark and charged lepton have a huge mass gap between the generations, and the pattern of the mixing in the quark and lepton should be explained by BSM. The flavor symmetry is one of the promising ideas to explain the flavor structure of the Yukawa coupling regardless of the Abelian or non-Abelian cases, and they are applied to not only the quark sector but also the lepton sector within the neutrino physics \cite{Maekawa:2001yh, Maekawa:2001uk, Dreiner:2003hw, Altarelli:2005yx, Kanemura:2007yy, Kamikado:2008jx, Araki:2008ek, King_2013, Dery:2016fyj, Gautam:2019pce, Higaki:2019ojq}. The Froggatt-Nielsen mechanism \cite{Froggatt:1978nt} can explain the mass hierarchy and the mixing pattern in the quark sector, so-called, the Cabibbo-Kobayashi-Maskawa (CKM) matrix, by introducing the additional $\mathrm{U}(1)$ symmetry and the SM gauge singlet field, "flavon". The Froggatt-Nielsen mechanism is applied to the SSM with the discrete symmetry as above mentioned \cite{Dreiner:2003yr, Dreiner:2006xw, Dreiner:2007vp}, where the discrete symmetry is embedded into the Abelian gauge symmetry which is identified with the flavor symmetry. Then, the flavor charge is strictly restricted by the gauge anomaly cancellation and requirement of the fermion mass hierarchy. 

The matter triality ($M_3$) which is introduced to prohibit the baryon number instead of the R-parity is one of the classes of the $\mathbb{Z}_3$ symmetry derived in \cite{Ibanez:1991hv, Ibanez:1991pr, Lee:2007qx}. The matter triality naturally requires the three-generation of the right-handed neutrino (RHN) to cancel the discrete anomaly, thus, the model predicts the additional state of the neutrino. The additional neutrino (so-called "sterile neutrino") only couples with the SM particle through the active neutrino, and the sterile neutrino plays various roles in the cosmology and particle physics depending on its mass scale \cite{Adhikari:2016bei}. If the mass of the lightest sterile neutrino is around ${\mathrm{keV}}$, it could be a candidate for the hot/warm dark matter \cite{Dodelson:1993je, Shi:1998km, Abazajian:2001nj, Shaposhnikov:2006xi, Asaka:2006rw, Asaka:2006nq, Abada:2014zra, Boyarsky:2018tvu, Abazajian:2021zui}. Otherwise, the $\mathrm{MeV}$-scale sterile neutrino might be related to the neutrinoless double beta decay violating the lepton number via the Majorana mass term \cite{Bilenky:2001rz, DellOro:2016tmg, Asaka:2020lsx, Asaka:2020wfo}.

 In the SSM with $M_3$, there is only allowed the $L$ violation term $\left[ {\hat{\nu}_{i}^c}{\hat{\nu}_{j}^c}{\hat{\nu}_{k}^c} \right]_F$, where $i,j,k$ are the generation index. This $L$ violating operator provides the Majorana mass term of RHN if the RH-sneutrino obtain the vacuum expectation value (VEV). As we will see, the seesaw mechanism in the SSM with the matter triality can be realized via the Majorana mass generated from the VEV of the RH-sneutrino which has the scale around the soft SUSY breaking mass $m_{\mathrm{soft}}$. 
Note that such $L$ violating operator appears in the "$\mu$ from $\nu$ SSM", where the supersymmetric Higgs mass is generated from the VEV of RH-sneutrino, and the neutrino mass spectrum and the mixing angle are derived \cite{Bratchikov:2005vp, Escudero:2008jg, Fidalgo:2009dm, Fidalgo:2009dm, LopezFogliani:2010bf, Zhang:2014osa, Mitsou:2015eka}. There are also the neutrino mass models in the RpV scenario, where the neutrino mass can be generated at the loop level, even if there is no tree-level Majorana mass of the right-handed neutrino \cite{Dreiner:2006xw, Dreiner:2007uj, Dreiner:2010ye, Dreiner:2011ft}.

In this research, we clarify the seesaw (SS) mechanism in the SSM with $M_3$ by using the $L$ violating operators. The matter triality is regarded as the remnant of the flavor symmetry after the spontaneous symmetry breaking. We assume that the flavor symmetry can be identified with the gauged $\mathrm{U}(1)_X$. 
Due to RpV, the mass matrix of the neutral fermion consist of the neutralino, LH-neutrino and RH-neutrino. 
Thus, the effective mass of the active neutrino is generated from the mixing with the two contributions, RHN which is associated with the Type-I seesaw mechanism and MSSM neutralino.
We analytically evaluate the expression of the active neutrino mass by using the seesaw approximation and derive the dependence in terms of the flavor charge of $\mathrm{U}(1)_X$. In addition to the flavor ansatz for the quark and the charged lepton, the mass spectrum of the active neutrino can determine the ansatz for the neutrino sector. Because of gauging the flavor symmetry, the flavor charge is strictly restricted in the parameter space since the symmetry must be anomaly-free. We consider the Green-Schwarz mechanism \cite{Green:1984sg} to cancel the mixed anomaly with the SM gauge groups. 
The model is motivated because of the existence of the light states of RHN. Although the Type-I seesaw mechanism includes RHN with the intermediate scale, such a heavier state of neutrino could not be detected directly. On the other hand, we show that the seesaw mechanism in the SSM with $M_3$ predicts the sterile neutrino with the mass below the soft SUSY breaking scale. Depending on the flavor charge, sterile neutrinos with masses between the keV-MeV scale are predicted, which can affect dark matter in cosmology and the observation of neutrinoless double beta decay \cite{PIENU:2011aa, PIENU:2017wbj, NA62:2020mcv, SHiP:2018xqw}. 

This paper is organized as follows. Firstly, in Sec.\ref{sec:M3 Model}, we introduce SSM with the matter triality. Then, Sec.\ref{sec:Flavor Model}, we review the Froggatt-Nielsen mechanism and provide the ansatz of the Yukawa coupling to explain the realistic quark and lepton mass hierarchy and mixing. In this section, the mass spectrum of the active neutrino is also derived by using the seesaw approximation as the extension of the Type-I seesaw mechanism. In the Sec.\ref{sec:Numerical Analysis}, we analyze the flavor charge satisfying the anomaly cancellation condition and the ansatz of Yukawa coupling. The concrete assignment of the flavor charge is given in the Appendix \ref{sec:Charge}. Then, we discuss the phenomenological aspect of this flavor model. By choosing the flavor charge, the model might include the light state of sterile neutrinos, and we can find that the anomaly-free flavor charge assignments are restricted by the neutrinoless double beta decay or the search for heavy neutral leptons. Finally, in Sec.\ref{sec:Conclusion}, we summarize this paper.

\section{Model}
\label{sec:M3 Model}
\subsection{Supersymmetric Standard Model with Matter triality}
 In Refs.\cite{Ibanez:1991hv,Ibanez:1991pr}, the alternative class of the R-parity has been proposed to prohibit the baryon number violation up to the dimension-5 operators. If we consider the anomaly cancellation of the discrete symmetry, the matter triality is allowed when the three-generation of right-handed neutrino are introduced. This discrete charge assignment is given in Tab.\ref{tb:M3charge}, where the superfield is rotated as $\hat{\Phi}\rightarrow e^{2\pi iq_\Phi /3}\hat{\Phi}$ under $M_3$.
\begin{table}[htb]
\centering
\caption{The $M_3$ charge assignment of each superfield.}
\begin{tabular}{c|cccccccccc}\hline\hline
Field & $\hat{Q}$ & $\hat{u}^c$& $\hat{d}^c$& $\hat{L}$& $\hat{e}^c$& $\hat{\nu}^c$ &$\hat{H}_u$ &$\hat{H}_d$\rule[-0pt]{0pt}{12pt} \\ \hline
$q_\Phi$ & $0$ & $2$& $1$& $1$& $0$& $1$ &$1$ &$2$ \rule[-0pt]{0pt}{12pt}\\ \hline\hline
\end{tabular}
\label{tb:M3charge}
\end{table}

 The superpotential in the SSM with $M_3$ is given by 
\begin{align}
W = & y^u_{ij}\hat{u}^c_i \hat{Q}_j \hat{H}_u+y^d_{ij}\hat{d}^c_i \hat{Q}_j \hat{H}_d+y^e_{ij}\hat{e}^c_i \hat{L}_j \hat{H}_d+y^\nu_{ij}\hat{\nu}^c_i \hat{L}_j \hat{H}_u+\frac{1}{3!}\kappa_{ijk}\hat{\nu}^c_i \hat{\nu}^c_j \hat{\nu}^c_k+\mu \hat{H}_u\hat{H}_d,
\label{eq:WinM3}
\end{align}
where $\hat{Q}_i = (\hat{u}_i, \hat{d}_i)$, $\hat{L}_i = (\hat{\nu}_i, \hat{e}_i)$, $\hat{H}_u^T = (\hat{H}_u^+, \hat{H}_u^0)$, and $\hat{H}_d^T = (\hat{H}_d^0, \hat{H}_u^-)$, and $i,j,k=1,2,3$ are the generation index of quarks and leptons. The color and weak indices for each fundamental representation is abbreviated and $y^{u,d,e,\nu}_{ij}$ are the Yukawa coupling of up-, down-type quark, charged lepton and neutrino, respectively. The only lepton number violating term $\kappa_{ijk}$ is totally symmetric for the permutation of the generation index. As we can see, the baryon number is preserved up to the dimension-5, thus the logevity of the proton is ensured instead of the R-parity. 

In the SUSY breaking sector, the Lagrangian by the soft SUSY breaking terms is also written down,
\begin{align}
-{\cal L}_{\mathrm{soft}}= & (m_{\tilde{Q}}^2)_{ij}\tilde{Q}^\ast_i\tilde{Q}_j + (m_{\tilde{u}}^2)_{ij}{\tilde{u}}_i^{c\ast}{\tilde{u}}_j^{c}+(m_{\tilde{d}}^2)_{ij}{\tilde{d}}_i^{c\ast}{\tilde{d}}_j^{c} + (m_{\tilde{L}}^2)_{ij}\tilde{L}^\ast_i\tilde{L}_j + (m_{\tilde{e}}^2)_{ij}{\tilde{e}}_i^{c\ast}{\tilde{e}}_j^{c} \nonumber \\
& +(m_{\tilde{\nu}}^2)_{ij}{\tilde{\nu}}_i^{c\ast}{\tilde{\nu}}_j^{c} + m_{H_u}^2H_u^{\ast}H_u + m_{H_d}^2H_d^{\ast}H_d + b[H_uH_d +c.c. ] \nonumber \\ 
& + \left[A^u_{ij}{\tilde{u}}_i^{c}\tilde{Q}_jH_u +A^d_{ij}{\tilde{d}}_i^{c}\tilde{Q}_jH_d + A^e_{ij}{\tilde{e}}_i^{c}\tilde{L}_jH_d + A^\nu_{ij}{\tilde{\nu}}_i^{c}\tilde{L}_jH_u +\frac{1}{3!}A^\kappa_{ijk}{\tilde{\nu}}_i^{c}{\tilde{\nu}}_j^{c}{\tilde{\nu}}_k^{c}+c.c. \right] \nonumber \\
&-\frac{1}{2}\left( M_3{\tilde{\lambda}}_3{\tilde{\lambda}_3}+M_2{\tilde{\lambda}}_2{\tilde{\lambda}_2}+M_1{\tilde{\lambda}}_1{\tilde{\lambda}_1}+ c.c. \right).
\end{align}

After the $F$- and $D$-terms are integrated out, the CP-even neutral scalar potential can be evaluated in terms of the parameters in the superpotential Eq.(\ref{eq:WinM3}). 
In general, the Higgs doublets and the sneutrinos can develop into the complex VEVs,
\begin{align}
\langle H_u^0 \rangle = v_u e^{i\phi_u} ,\,\,\, \langle H_d^0 \rangle = v_d e^{i\phi_d} ,\,\,\, \langle {\tilde{\nu}}_i \rangle = v_{L_i}e^{i\phi_{L_i}} ,\,\,\, \langle {\tilde{\nu}}_i^{c} \rangle =v_{R_i}e^{i\phi_{R_i}} .\,\,\,
\end{align}
Let us assume that the parameters in the superpotential are real. Then the neutral scalar potential can be written in terms of the complex VEVs at the tree level,
\begin{align}
V = & V_F + V_D + V_{\mathrm{soft}} ,
\end{align}
where 
\begin{align}
V_D = & \frac{1}{8}\left(g^2+g'^2)(v_{L_i}v_{L_i} + v_u^2+v_d^2\right)^2 \\
V_F = & \mu^2v_d^2 + 2\mu v_d v_{R_i}y_{ij}^\nu v_{L_j} \cos\left(\phi_v-\chi_j+\phi_{R_i} \right)\nonumber \\
& + (v_{R_i}y_{ij}^\nu v_{L_j})(v_{R_l}y_{lm}^\nu v_{L_m})\cos\left(\chi_j-\chi_m-\phi_{R_i}+\phi_{R_l}\right)\nonumber \\
& + \mu^2v_u^2 + v_u^2v_{R_i}(y^\nu y^{\nu T})_{ij} v_{R_j}\cos\left(\phi_{R_i} -\phi_{R_j}\right) +v_u^2v_{L_i}(y^{\nu T}y^\nu)_{ij}v_{L_j} \cos\left({\chi_{i}-\chi_j}\right) \nonumber \\
& + v_uv_{L_m}y^{\nu T}_{mi}\kappa_{ijk}v_{R_j}v_{R_k}\cos\left(\chi_m+\phi_{R_j}+\phi_{R_k}\right)\nonumber \\ & +\frac{1}{4}\kappa_{ijk}\kappa_{ilm}v_{R_j}v_{R_k}v_{R_l}v_{R_m}\cos\left(\phi_{R_j}+\phi_{R_k}-\phi_{R_l}-\phi_{R_m}\right) \\
V_{\mathrm{soft}} = &  m_{H_u}^2v_u^2 + m_{H_d}^2v_d^2 + (m_{\tilde{L}}^2)_{ij}v_{L_i}v_{L_j}\cos\left(\chi_i-\chi_j\right)+(m_{\tilde{{\nu}}}^2)_{ij}v_{R_i}v_{R_j}\cos\left( \phi_{R_i}-\phi_{R_j}\right)  \nonumber \\ 
& + bv_uv_d \cos\phi_v + 2A^\nu_{ij}v_{R_i}v_{L_j}v_u\cos\left( \chi_j-\phi_{R_i}\right) +\frac{1}{3}A^\kappa_{ijk}v_{R_i}v_{R_j}v_{R_k}\cos\left(\phi_{R_i}+\phi_{R_j}+\phi_{R_k} \right) .
\end{align}
The phase in the potential can be defined as 
\begin{align}
\phi_v= \phi_u+\phi_d ,\,\,\, \chi_i=\phi_{L_i}+\phi_u.
\end{align}
The potential can be found that it depends on the eight real VEVs $(v_u,v_d,v_{L_i},v_{R_j})$ and the seven phases $(\phi_v, \chi_i, \phi_{R_j})$, thus, we obtain the fifteen minimization condition. In general, the derivation of the global minimum in SSM without R-parity is a complex problem, the analysis is beyond the scope of this paper. However, the VEV of the RH-sneutrino determines the scale of the Majorana mass, thus we can estimate the scale of $v_{R_i}$.

The VEVs of the LH/RH sneutrino are determined via the minimization conditions.
\begin{align}
t_{v_{I}}^0 = \dfrac{\partial V}{\partial v_I}=0, \,\,\,\, v_I=v_u,v_d,v_{L_i},v_{R_i},\phi_v, \chi_i, \phi_{R_i}.
\end{align}
As we will see, the neutrino Yukawa coupling should be suppressed $y^\nu_{ij}<10^{-7}$ to explain the correct neutrino mass scale. Then, we can neglect the Yukawa coupling in the minimization condition. This leads to the small VEVs for the LH-sneutrino in comparison with the EW scale in order to keep the vacuum of the electroweak symmetry breaking (EWSB), $v=\sqrt{v_{L_i}^2+v_u^2+v_d^2}\sim \sqrt{v_u^2+v_d^2}$.  Then, the minimization equations of the RH-sneutrino are given at the limit $y^\nu\rightarrow 0$ and $v_{L_i}\rightarrow 0$,
\begin{align}
 t_{R_i}^0 \sim & \kappa_{nij}\kappa_{nlm}v_{R_j}v_{R_l}v_{R_m}\cos(\phi_{R_i}+\phi_{R_j}-\phi_{R_l}-\phi_{R_m}) \nonumber \\
 & + 2(m_{\tilde{\nu}}^2)_{ij}v_{R_j}\cos(\phi_{R_i}-\phi_{R_j}) + \frac{1}{3}A^{\kappa}_{ijk}v_{R_j}v_{R_k}\cos(\phi_{R_i}+\phi_{R_j}+\phi_{R_k}) = 0. 
\end{align}
This equation implies that the RH-sneutrino develops into the VEV around soft mass scale $m_{\mathrm{soft}}$ which is determined by the mediation mechanism of the SUSY breaking. Thus, in the following discussion, we assume that the RH-sneutrino acquires the VEVs nearby $m_{\mathrm{soft}}$ while the VEVs of the LH-sneutrino are neglected, $\braket{\tilde{\nu}_{L_i}}=0$.
In the next section, we will show the roll of the VEV of RH-sneutrino in the seesaw mechanism. 

\subsection{Seesaw Mechanism in the SSM with $M_3$}
In our model, the active neutrinos observed in the neutrino oscillation experiments \cite{Zyla:2020zbs,Esteban:2020cvm} consist of the mixture of the left-handed neutrinos, the right-handed neutrinos, and the MSSM neutralinos due to the R-parity violation. The Majorana mass term of the right-handed neutrino is induced by the lepton number violation term, $\kappa_{ijk}$ and the RH-sneutrino VEV.

In our notation, the fermionic component in the chiral superfield is described by the left-handed Weyl fermion, thus the LH/RH-neutrino are denoted as $\nu_i = \nu_{L_i}$ and ${\nu}_i^{c} = \nu_{R_i}^\dagger$. We can also reduce the superpotential into the Lagrangian in terms of the component fields,
\begin{align}
{\cal L}_\nu =& -y^\nu_{ij}{\bar{\nu}_{R_i}} {\nu_{L_j}}H_u^0 -y^\nu_{ij}\widetilde{\nu}_{R_i}^c {\nu_{L_j}}\widetilde{H}_u^0 -y^\nu_{ij}\bar{\nu}_{R_i} \widetilde{\nu}_{L_j}\widetilde{H}_u^0 + c.c. \nonumber \\
& \,\, +y^\nu_{ij}{\bar{\nu}_{R_i}} {e_{L_j}}H_u^+ + y^\nu_{ij}\widetilde{\nu}_{R_i}^c {e_{L_j}}\widetilde{H}_u^+ +y^\nu_{ij}\bar{\nu}_{R_i} \widetilde{e}_{L_j}\widetilde{H}_u^+ + c.c. \nonumber \\
& \,\, + \frac{1}{2}\left( \kappa_{ijk}\tilde{\nu}^{c}_{R_i}\bar{\nu}_{R_j} \nu_{R_k}^c+ \kappa_{ijk}^\ast\tilde{{\nu}}_{R_i}^{c\ast} \bar{\nu}_{R_j}^c \nu_{R_k}\right).
\label{eq:Lnu}
\end{align}
In Eq.(\ref{eq:Lnu}), we obtain the Dirac mass term and the Majorana mass term for the neutrino due to the VEV of up-type Higgs boson and RH-sneutrinos:
\begin{align}
& y^\nu_{ij}{\bar{\nu}_{R_i}} {\nu_{L_j}}H_u^0 \rightarrow \langle H_u^0 \rangle y^\nu_{ij}{\bar{\nu}_{R_i}} {\nu_{L_j}}=\left( m_D \right)_{ij}{\bar{\nu}_{R_i}} {\nu_{L_j}} \\
& \kappa_{ijk}\tilde{\nu}^{c}_{R_i}\bar{\nu}_{R_j} \nu_{R_k}^c \rightarrow \kappa_{ijk}\langle {\tilde{\nu}}_{R_i}^c\rangle\bar{\nu}_{R_j} \nu_{R_k}^c = \left(M_R \right)_{ij}\bar{\nu}_{R_i} \nu_{R_j}^c.
\end{align}

In addition to the neutrino mass matrix, due to the R-parity violation, the neutrinos are mixed with the MSSM neutralino in the neutral fermion mass matrix. In the basis $\Psi_0^T=(\tilde{\lambda}_1, \tilde{\lambda}_2, \tilde{H_u^0}, \tilde{H_d^0},\nu_{L_i}, \nu_{R_i}^c)$, the neutralino-neutrino mass matrix is given by the $10\times10$ matrix if the smallness of the VEV of LH-sneutrino is assumed, 

{\footnotesize
\begin{align}
&{\cal{M}} =\nonumber \\
&\begin{pmatrix}
M_1 & 0 & -A\langle H_d^0\rangle^\ast & A\langle H_u^0\rangle^\ast & 0 & 0 & 0 & 0 & 0 & 0 \\
0 & M_2 & B\langle H_d^0\rangle^\ast & -B\langle H_u^0\rangle^\ast & 0 & 0 & 0 & 0 & 0 & 0 \\
-A\langle H_d^0\rangle^\ast & B\langle H_d^0\rangle^\ast & 0 & \mu & 0 & 0 & 0 & 0 & 0 & 0 \\
A\langle H_u^0\rangle^\ast & -B\langle H_u^0\rangle^\ast & \mu & 0 &  y^\nu_{k1}\langle {\tilde{\nu}}^c_k\rangle & y^\nu_{k2}\langle {\tilde{\nu}}^c_k\rangle & y^\nu_{k3}\langle {\tilde{\nu}}^c_k\rangle & 0 & 0 & 0 \\
0 & 0 & 0 & y_{k1}^\nu \langle {\tilde{\nu}}^c_k\rangle & 0 & 0 & 0 & y_{11}^\nu \langle H_u^0\rangle & y_{21}^\nu \langle H_u^0\rangle & y_{31}^\nu \langle H_u^0\rangle \\
0 & 0 & 0 & y_{k2}^\nu \langle {\tilde{\nu}}^c_k\rangle & 0 & 0 & 0 & y_{12}^\nu \langle H_u^0\rangle & y_{22}^\nu\langle H_u^0\rangle & y_{32}^\nu\langle H_u^0\rangle\\
0 & 0 & 0 & y_{k3}^\nu \langle {\tilde{\nu}}^c_k\rangle & 0 & 0 & 0 & y_{13}^\nu \langle H_u^0\rangle & y_{23}^\nu \langle H_u^0\rangle& y_{33}^\nu \langle H_u^0\rangle \\
0 & 0 & 0 & 0 & y_{11}^\nu \langle H_u^0\rangle & y_{12}^\nu \langle H_u^0\rangle & y_{13}^\nu \langle H_u^0\rangle & \kappa_{11k} \langle {\tilde{\nu}}^c_k\rangle & \kappa_{12k} \langle {\tilde{\nu}}^c_k\rangle& \kappa_{13k} \langle {\tilde{\nu}}^c_k\rangle \\
0 & 0 & 0 & 0 & y_{21}^\nu\langle H_u^0\rangle & y_{22}^\nu \langle H_u^0\rangle & y_{23}^\nu \langle H_u^0\rangle & \kappa_{21k} \langle {\tilde{\nu}}^c_k\rangle & \kappa_{22k} \langle {\tilde{\nu}}^c_k\rangle & \kappa_{23k} \langle {\tilde{\nu}}^c_k\rangle \\
0 & 0 & 0 & 0 & y_{31}^\nu \langle H_u^0\rangle & y_{32}^\nu\langle H_u^0\rangle & y_{33}^\nu \langle H_u^0\rangle & \kappa_{31k} \langle {\tilde{\nu}}^c_k\rangle & \kappa_{32k} \langle {\tilde{\nu}}^c_k\rangle & \kappa_{33k} \langle {\tilde{\nu}}^c_k\rangle
\end{pmatrix},
\end{align} 
}
where $A=\sqrt{(g^2+g'^2)/2}\sin\theta_w$ and $B=\sqrt{(g^2+g'^2)/2}\cos\theta_w$.
The neutralino-neutrino mass matrix can be evaluated by the block diagonalization. Let us denote the MSSM neutralino as $\psi_n^T=(\tilde{\lambda}_1, \tilde{\lambda}_2, \tilde{H_d^0}, \tilde{H_u^0})$,
\begin{align}
\frac{1}{2} \bar{\Psi} {\cal M} \Psi + h.c.
= \frac{1}{2}
\begin{pmatrix}
\overline{\psi_n^c} & \overline{\nu_L^c} & \overline{\nu_R}
\end{pmatrix}
\begin{pmatrix}
M_N & M_X & 0 \\
M_X^T & 0 & m_D^T \\
0 & m_D & M_R
\end{pmatrix}
\begin{pmatrix}
\psi_n\\
\nu_L\\
\nu_R^c 
\end{pmatrix} + h.c.
\end{align}
The mixing between the neutralino and the LH-neutrino are given by the $3\times 4$ matrix $M_X^T$. In general, the complex symmetric matrix is diagonalized by the unitary matrix and its transposed one. Thus, $\Psi$ is rotated by the unitary matrices $U_{LR}, U_{NL},$ and $U_{RN}$, which are the block-diagonalizing matrix between $(\nu_L, \nu_R^c)$, $(\psi_n, \nu_L)$ and $(\psi_n, \nu_R^c)$, respectively.
\begin{align}
\begin{pmatrix}
\psi_n\\
\nu_L\\
\nu_R^c 
\end{pmatrix}= U_{LR}U_{NL}U_{RN}
\begin{pmatrix}
U_\chi & 0 & 0 \\
0 & U_\nu & 0 \\
0 & 0 & U_N \\
\end{pmatrix}
\begin{pmatrix}
\chi \\
\nu \\
N
\end{pmatrix},
\end{align}
where the unitary matrices $U_\chi$, $U_\nu$, and $U_N$ diagonalize each block matrix.
Then, the mass matrix ${\cal M}$ can be block-diagonalized by using the seesaw approximation if $m_D \ll M_R$ and $M_X \ll M_N$, 
\begin{align}
\therefore \frac{1}{2} \bar{\Psi} {\cal M} \Psi + h.c. 
&= \Psi_m^T \left( U_{LR}U_{NL}U_{RN} \right)^T {\cal M} U_{LR}U_{NL}U_{RN} \Psi_m\\
& \sim 
\begin{pmatrix}
\bar{\chi} & \bar{\nu} & \bar{N}
\end{pmatrix}
\begin{pmatrix}
U_\chi^TM_{\chi}U_\chi & 0 & 0 \\
0 & U_\nu^Tm_{\mathrm{eff}} U_\nu & 0 \\
0 & 0 & U_N^T M_N U_N 
\end{pmatrix}
\begin{pmatrix}
\chi \\
\nu \\
N
\end{pmatrix}
\end{align}
After the seesaw approximation, we obtain the approximate form of the mass matrix for neutralino and neutrinos:
\begin{align}
M_\chi &= M_N + \frac{1}{2}\left( \varepsilon\varepsilon^\dagger M_N^T -M_N \varepsilon^\ast\varepsilon^T \right) \\
m_{\mathrm{eff}} &= -m_D^T M_R^{-1}m_D - M_X^TM_N^{-1}M_X\\
M_N &= M_R + \frac{1}{2}\left( \theta^\dagger\theta M_R + M_R^T\theta^T\theta^\ast \right)
\end{align}
We can see that the effective mass matrix of the active neutrino is obtained via the approximation like the Type-I SS mechanism.
These unitary matrices are approximately expressed by the mixing angles,
\begin{align}
U_{LR} = &
\begin{pmatrix}
{\mathbb{I}} & 0 & 0 \\
0 & {\mathbb{I}}-\frac{1}{2}\theta^\ast\theta^T & \theta^\ast \\
0 & -\theta^T & {\mathbb{I}}-\frac{1}{2}\theta^T\theta^\ast
\end{pmatrix} , \\
U_{NL} = &
\begin{pmatrix}
{\mathbb{I}}-\frac{1}{2}\varepsilon \varepsilon^\dagger & \varepsilon & 0 \\
-\varepsilon^\dagger & {\mathbb{I}}-\frac{1}{2}\varepsilon^\dagger \varepsilon & 0 \\
0 & 0 & {\mathbb{I}} 
\end{pmatrix},\\
U_{RN} = &
\begin{pmatrix} 
{\mathbb{I}} & 0 & -\frac{1}{2}\eta \\
0 & {\mathbb{I}} & 0 \\
-\frac{1}{2}\xi & 0 & {\mathbb{I}}
\end{pmatrix},
\end{align}
where the mixing angles between each neutral fermion are written as 
\begin{align}
\theta &=m_D^TM_R^{-1} \\
\varepsilon &=-M_N^{-1}M_X \\
\eta &=M_N^{-1}M_X\theta^\ast \\
\xi &= M_R^{-1}\theta^\dagger M_X^T.
\end{align}
Note that the mixing angle should be suppressed to carry out the seesaw approximation\footnote{In the seesaw approximation, the specific combination of the mixing angle ${M_X}\varepsilon^\ast \varepsilon^T$ may excess $M_X^T M_N^{-1}M_X$. In the numerical analysis, we assign the condition so that the seesaw approximation is valid.}. The concrete form of the mixing angle is given in the Appendix.\ref{sec:mixing}.
The procedure of the seesaw approximation indicates that we have the sterile state of the neutrino (sterile neutrino) which couples with the active neutrino via Yukawa coupling. By using the seesaw formula, the interaction with the sterile neutrino $N_I$ arise from the gauge interaction terms,
\begin{align}
{\cal L}_{\mathrm{int}} \supset &\,\, \frac{g}{2c_w}N_I^\dagger\Theta^\dagger_{Ii}\overline{\sigma^\mu}\nu_{L_i}Z_\mu +\frac{g}{2c_w}\nu_{L_i}^\dagger\overline{\sigma^\mu}\Theta_{iI}N_I Z_\mu+\frac{g}{2}N_I^\dagger\Theta^\dagger_{Ii}\overline{\sigma^\mu}e_{i}W^+_\mu + \frac{g}{2}e_{i}^\dagger\overline{\sigma^\mu}\Theta_{iI}N_IW^-_\mu\nonumber \\
& \,\,\, +\frac{g}{2\sqrt{2}s_\beta m_W}\left(M_R\theta^TU_N\right)_{Ij}n_{u1}N_I^T\nu_{L_j}h^0 + \frac{g}{2\sqrt{2}s_\beta m_W}\left(M_R\theta^TU_N\right)_{Ij}c_{u1}N_I^Te_{j}h^+ +h.c, 
\label{eq:Lint}
\end{align}
where $g$ is the $\mathrm{SU}(2)_L$ gauge coupling constant. The mixing between the active and the sterile neutrino is given by $\Theta=\theta^\ast U_N$ and the rotation matrices for the neutral and charged scalar is defined as $n_{u\alpha}$ and $c_{u\alpha}$. The coupling strength between the sterile neutrino and active neutrino is determined by the mixing $\theta$ which can be rewritten in terms of the flavor charge by assigning the Froggatt-Nielsen mechanism.

\section{Froggatt-Nielsen Mechanism and Flavor Ansatz}
\label{sec:Flavor Model}
In this section, we will consider the Froggatt-Nielsen mechanism \cite{Froggatt:1978nt} which explains the hierarchical structure of quark/lepton masses. In this context, we combine with the matter triality to simultaneously explain the baryon number preserving. Furthermore, $M_3$ can be originated from the gauged $\mathrm{U}(1)_X$ symmetry, and let us identify it with the Froggatt-Nielsen $\mathrm{U}(1)_X$ symmetry. By this construction, the flavor $\mathrm{U}(1)_X$ charges are not only constrained from the ansatz that realizes the mass and mixing angle but also the anomaly cancellation condition with the mixed anomaly of the SM gauge groups. 

Firstly, in the Froggatt-Nielsen mechanism, the additional gauge singlet superfield, flavon $\hat{S}$ is introduced and the Yukawa coupling is filled up by the power of the flavon since it is no longer invariant under the gauge symmetry. Therefore, the superpotential under the gauged $\mathrm{U(1)}_X$ symmetry are written with the ${\cal O}(1)$ coefficients,
\begin{align}
W = & g^u_{ij}\hat{u}^c_i \hat{Q}_j \hat{H}_u\left( \dfrac{\hat{S}}{M_{\mathrm{Pl}}}\right)^{n^u_{ij}}+g^d_{ij}\hat{d}^c_i \hat{Q}_j \hat{H}_d\left( \dfrac{\hat{S}}{M_{\mathrm{Pl}}}\right)^{n^d_{ij}}+g^e_{ij}\hat{e}^c_i \hat{L}_j \hat{H}_d\left( \dfrac{\hat{S}}{M_{\mathrm{Pl}}}\right)^{n^e_{ij}}\nonumber \\
&+g^\nu_{ij}\hat{\nu}^c_i \hat{L}_j \hat{H}_u\left( \frac{\hat{S}}{M_{\mathrm{Pl}}}\right)^{n^\nu_{ij}} +g^\kappa_{ijk}\hat{\nu}^c_i\hat{\nu}^c_j\hat{\nu}^c_k\left( \frac{\hat{S}}{M_{\mathrm{Pl}}}\right)^{n^\kappa_{ijk}}+\mu g^\mu \hat{H}_u\hat{H}_d\left( \frac{\hat{S}}{M_{\mathrm{Pl}}}\right)^{n^\mu},
\label{eq:WinfM3}
\end{align}
where $g_{ij}^{u,d,e,\nu}, g^\kappa_{ijk}$, and $g^\mu$ are the ${\cal O}(1)$ coefficients for each coupling.
The power is determined by the flavor charge, for example, $n_{ij}^u=X_{U_i}+X_{Q_j}+X_{H_u}$. After the flavor symmetry breaking, the flavon acquires the VEV around $M_{\mathrm{Pl}}$, this ratio is described by $\epsilon= \langle S\rangle/M_{\mathrm{Pl}}$ and it should be nearby the Cabbibo angle $\epsilon\sim 0.22$. The hierarchical structure of the Yukawa coupling can be controlled by the flavor symmetry.
The matter triality is preserved as the remnant symmetry after the flavor symmetry breaking, if we choose the flavor charge $X_{\Phi_i}=q_{\Phi}+3k_{\Phi_i}$, where $X_{\Phi_i}$, $q_{\Phi}$ and $k_{\Phi_i}$ are the flavor charge of the gauged $\mathrm{U}(1)_X$ symmetry, the charge of the matter triality and the parameter which shows the dependence of the generation, respectively. Note that the generation dependence of the flavor symmetry comes from this parameter $k_{\Phi_i}$. 

\subsection{Quark and Charged Lepton Sector}
First of all, let us consider the Yukawa ansatz for the quark and lepton. In Ref.\cite{Dreiner:2006xw, Hinata:2020cdt}, the ansatz of the Yukawa matrix for the up-type quark is given by
\begin{equation}
n_{ij}^u =
\begin{pmatrix}
8 & 7-y & 5-y \\
5+y & 4 & 2 \\
3+y & 2 & 0
\end{pmatrix}_{ij},
\label{eq:YUansatz}
\end{equation}
where $y=0,1$ is an integer parameter related to the CKM mixing matrix. For the down-type quark and charged electron, those ansatzs can be also written in terms of some parameters,
\begin{equation}
n_{ij}^d =
\begin{pmatrix}
4+x & 3-y+x & 1-y+x \\
3+y+x & 2+x & x \\
3+y+x & 2+x & x
\end{pmatrix}_{ij},
\label{eq:YDansatz}
\end{equation}
\begin{equation}
n_{ij}^e=
\mathrm{diag}\left(4+z+x,\ 2+x,\ x \right)_{ij},
\label{eq:YEansatz}
\end{equation}
where $x=0,1,2,3$ and $ z=0,1$. The integer parameter $x$ can be interpreted as the ambiguity of $\tan\beta\sim v_u/v_d$, {\it i.e.}, $m_b/m_t \sim \epsilon^x \cot\beta = \epsilon^x v_d/v_u$. These ansatzs reproduce the observed mass hierarchy of the quark and charged lepton \cite{Zyla:2020zbs},
\begin{align}
& \frac{m_u}{m_t}: \frac{m_c}{m_t} = \epsilon^8: \epsilon^4 \\
& \frac{m_d}{m_b}: \frac{m_s}{m_b} = \epsilon^4: \epsilon^2 \\
& \frac{m_e}{m_\tau}: \frac{m_\mu}{m_\tau} = \epsilon^{4+z}: \epsilon^2 .
\end{align}
On the other hands, derived from the matrix structure, the CKM mixing matrix can be obtained by the parameter $y$ \cite{Harnik:2004yp},
\begin{equation}
V_{\mathrm{CKM}} \sim
\begin{pmatrix}
1 & \epsilon^{1+y} & \epsilon^{3+y} \\
\epsilon^{1+y} & 1 & \epsilon^2 \\
\epsilon^{3+y} & \epsilon^2 & 1
\end{pmatrix}.
\end{equation}

\subsection{Higgs Sector}
\label{sec:Higgs sector}
The extended Higgs sector is introduced in the context of the string compactification, for example, orbifold model \cite{Font:1989aj} or magnetized orbifold model \cite{Aldazabal:2000sa}. On the other hand, the phenomenological aspect of the multiple Higgs doublet model has been discussed \cite{McWilliams:1980kj, Shanker:1981mj, Branco:2011iw, Aranda:2000zf}. Furthermore, if the Higgs doublets are charged under the extra gauge symmetry, they contribute to the anomaly cancellation in the supersymmetric extension \cite{Hinata:2020cdt}. In the previous work, we introduce the extra-Higgs doublets and analyze the contribution to the anomaly cancellation condition. These extra-Higgs doublets have to be decoupled from the theory at the high energy scale because such additional states of the scalar are constrained by the flavor changing neutral currents in the $K$, $B$, and $D$ mesons via the exchange of the neutral scalar \cite{Escudero:2005hk}.

We introduce the extra-Higgs doublets to consider the anomaly cancellation conditions. Let us denote the extra-Higgs doublets as $\hat{H}_{ua}$ and $\hat{H}_{da}$, where the representation of the gauge symmetry $\mathrm{SU}(2)_L\times \mathrm{U}(1)_Y\times\mathrm{U}(1)_X$ is defined as
\begin{align}
\hat{H}_{ua} :& \, \, \left({\bf 2}, \,\, +\frac{1}{2}, \,\, 2+3k_{H_{ua}} \right) \\
\hat{H}_{da} :& \, \, \left({\bf 2}, \,\, -\frac{1}{2}, \,\, 1+3k_{H_{da}} \right),
\end{align}
where the roman indices imply the generation of extra-Higgs fields $a=2,\cdots, N_h$. The flavor charge is uniquely determined to avoid the baryon number violation, and to prohibit the Yukawa coupling of quarks and leptons \cite{Hinata:2020cdt}.

The mass of the Higgs doublet can be generated by the supersymmetric mass term $\mu$ in the superpotential, which should be around the cutoff scale of the theory. On the other hands, if we introduce the non-minimal coupling in the K\"alher potential, the mass of the Higgs doublets which determines the scale of EWSB, associates with the SUSY breaking scale \cite{Giudice:1988yz}. Let us denote the generation indecies of the Higgs fields within the extra generation as Greek index $\alpha,\beta=1,\cdots, N_h$. Then, we assume that the non-minimal coupling in the K\"ahler potential, which induces the effectively supersymmetric Higgs mass,
\begin{align}
\tilde{g}_{\alpha\beta}^{\mu}\int d^4\theta \frac{\hat{\bar{Z}}}{M_{\mathrm{Pl}}} \biggl\{\Theta[-\Omega_{\alpha\beta}^\mu] \left(\frac{\hat{\bar{S}}}{M_{\mathrm{Pl}}} \right)^{-\Omega_{\alpha\beta}^\mu} +
\Theta[\Omega_{\alpha\beta}^\mu]\left(\frac{\hat{S}}{M_{\mathrm{Pl}}} \right)^{\Omega_{\alpha\beta}^\mu} \bigg\}\hat{H}_{u\alpha}\hat{H}_{d\beta} + c.c.,
\end{align}
where $\Omega_{\alpha\beta} = X_{H_{u\alpha}}+X_{H_{d\beta}}$ and $\Theta\left[ x\right]=1$ for $x\geq 0$ while $\Theta\left[ x\right]=0$ for $x< 0$. If we assume the gravity mediated SUSY breaking, then the F-term of the hidden sector superfields $\braket{F_{\bar{Z}}}=m_{\mathrm{soft}}M_{\mathrm{Pl}}$, where $m_{\mathrm{soft}}$ is soft SUSY breaking mass scale whose scale depends on the power of the flavon. After $\bar{Z}$ is integrated out, the "$\mu$-matrix" which relates to the soft mass is given by,
\begin{equation}
\mu_{\alpha\beta} = \tilde{g}_{\alpha\beta}^\mu m_{\mathrm{soft}}\epsilon^{|\Omega_{\alpha\beta}|} .
\end{equation}
Therefore, the effective $\mu$-matrix can be derived as
\begin{eqnarray}
\mu_{\alpha\beta} = {g}_{\alpha\beta}^\mu M_{\mathrm{Pl}} \Theta\left[\Omega_{\alpha\beta}\right]\epsilon^{\Omega_{\alpha\beta}} +\tilde{g}_{\alpha\beta}^\mu m_{\mathrm{soft}} \epsilon^{|\Omega_{\alpha\beta}|} .
\label{eq:muMAT}
\end{eqnarray}
If we choose the sign of the flavor charge, the decoupling can be realized because of the holomorphy of the superpotential. The MSSM Higgs fields should be around $100\mathrm{GeV}$ below the soft SUSY breaking scale to realize the EWSB, while the extra doublets can be heavy when the mass term is allowed in the superpotential. 
In the following discussion, let us identify the first pair of the Higgs doublet as the MSSM Higgs field, and the sum of the flavor charge is defined $w=\Omega_{11}$. Then, the soft SUSY breaking mass term connects to the EW scale,
\begin{equation}
\mu \sim \epsilon^w m_{\mathrm{soft}}.
\end{equation}
We refer to $k_{H_{u}}$ and $k_{H_{d}}$ as the charge of the first pair of Higgs doublets without any indication.

\subsection{Neutrino Sector}
To realize the observables in the neutrino sector, let us evaluate the flavor ansatz in the neutrino sector.
After block diagonalization, we can evaluate the light active neutrino mass matrix $m_{\mathrm{eff}}$:
\begin{align}
 m_{\mathrm{eff}} &= -M_X^T M_N^{-1} M_X - m_D^T M_R^{-1} m_D.
 \label{eq:meff}
\end{align}
The first term comes from the neutralino mass matrix, and we can evaluate the approximated expression of the contribution from the neutralino,
\begin{align}
-(M_X^T M_N^{-1} M_X)_{ij}& \sim -y_{ki}^\nu v_{R_k} \dfrac{(A^2M_2+B^2M_1)v_d^2}{\mu(2A^2M_2v_uv_d+B^2M_1v_d(v_u+v_d)+M_1M_2\mu)} y^\nu_{lj}v_{R_l},
\label{eq:MNtomeff}
\end{align}
where the phases of the VEV are neglected because those are irrelevant to the discussion of the mass scale. The neutralino contribution predicts the massless neutrino, because the mass matrix Eq.(\ref{eq:MNtomeff}) is not regular. However, the active neutrino is provided by Eq.(\ref{eq:MNtomeff}), if it dominant in Eq.(\ref{eq:meff}) and the next to the heaviest mode is generated from contribution from the RHN. In order to evaluate the contribution, we rewrite it in terms of the Cabbibo angle $\epsilon$ and the flavor charge. Furthermore, we can rewrite the soft mass $m_{\mathrm{soft}}$, $\tan\beta$ and the neutrino Yukawa coupling $y^\nu$ by using the flavor charge
\begin{align}
&m_{\mathrm{soft}} = v \epsilon^{-w} \\
&\tan\beta = \dfrac{m_t}{m_b}\epsilon^x \\
&y^\nu_{ij} \sim \epsilon^{k_{N_i}+k_{L_j}+k_{H_u}}
\end{align}
where $m_t$ and $m_b$ are the top and bottom quark mass, and the soft SUSY breaking mass $m_{\mathrm{soft}}$ is associated with the EW scale $v$ via parameter $w$ (See also Sec.\ref{sec:Higgs sector}). After the calculation, we can derive the contribution from the neutralino mass matrix in terms of the flavor charge,
\begin{align}
-\left(M_X^T M_N^{-1} M_X \right)_{ij}\sim \dfrac{v}{1+\tan^2\beta}\epsilon^{2(k_{H_u}+k_{N_3})-w+k_{L_i}+k_{L_j}},
\end{align}
In this expression, we have assumed the order of flavor charge $k_{N_1}\geq k_{N_2}\geq k_{N_3}$. We can find that the hierarchical structure in the effective mass only depends on the flavor charge of the lepton doublet $k_{L_i}$, however, the scale is related to the other parameters $w$ and $k_{N_3}$. 
On the other hands, the second term in the Eq.(\ref{eq:meff}) is originated from the Majorana mass matrix of the right-handed neutrino, which is same with the usual Type-I seesaw mechanism. The expression by the flavor charge is also evaluated:
\begin{align}
-\left(m_D^TM_R^{-1}m_D\right)_{ij} \sim v\sin^2\beta\epsilon^{2+2k_{H_u}+w+k_{L_i}+k_{L_j}-k_{N_3}}.
\end{align}
Again, the hierarchical texture comes from the flavor charge of the lepton doublet $k_{L_i}$, and the scale depends on the other parameters $w$ and $k_{N_3}$. The dependence of the flavor charge for the matrix texture does not change in the case of the Type-I seesaw mechanism, while the mass scale of the active neutrino is modified since the Majorana mass is generated via the VEV of the RH-sneutrino.
To compare the contribution from RHNs with the one from the neutralino, let us define the ratio between the two contributions,
\begin{align}
R= \dfrac{\left(M_X^T M_N^{-1} M_X \right)_{ij}}{\left(m_D^TM_R^{-1}m_D\right)_{ij}} = \cot^2\beta\epsilon^{-2w+3k_{N_3}-2}. \label{eq:rSS}
\end{align}
This means that the dominant SS contribution depends on the parameter $x, w$ and $k_{N_3}$. Note that, if we consider the low-scale SUSY breaking ($w\rightarrow 1$), then the VEV of the RH-sneutrino also decreases. This leads to the approximate realization of the Type-I seesaw mechanism because the mixing between the neutralino and LH-neutrino vanishes. 
Again, the next to the heaviest mode of the active neutrino have to be generated from the RHNs contribution in the neutralino dominated case. In this case, the ratio $R$ must be restricted to realize the correct neutrino data. We constrain the flavor charge so that the effective mass $m_{\mathrm{eff}}$ have the correct mass hierarchy, $R \geq \epsilon^3$.

By using the formula Eq.(\ref{eq:rSS}), we can derive the correct ansatz of the neutrino mass matrix for each case, depending on the neutrino mass hierarchy. The observed data indicates the two pattern of the hierarchical spectrum between the mass differences, $\Delta m_{ij}^2\equiv m_i^2-m_j^2$ \cite{Zyla:2020zbs,Esteban:2020cvm}.
\begin{description}
\item[Normal Hierarchical Spectrum (NH):] $m_1\ll m_2 < m_3$,
\begin{align}
m_2 &\sim \sqrt{\Delta m_{21}^2} \sim 8.6\times 10^{-3}\mathrm{eV},\nonumber \\
m_3 &\sim \sqrt{\Delta m_{21}^2+\Delta m_{32}^2} \sim 0.05 \mathrm{eV}.\nonumber  
\end{align}
\item[Inverted Hierarchical Spectrum (IH):] $m_3 \ll m_1< m_2$,
\begin{align}
m_1 &\sim \sqrt{|\Delta m_{21}^2+\Delta m_{32}^2|} \sim 0.0492\mathrm{eV},\nonumber \\
m_2 &\sim \sqrt{\Delta m_{21}^2} \sim 0.05 \mathrm{eV}. \nonumber 
\end{align}
\end{description}

For the evaluation of the mass spectrum of active neutrino and the leptonic mixing matrix, {\it i.e.,} Pontecorvo-Maki-Nakagawa-Sakata (PMNS) matrix, we derive the ansatz of the flavor charge for the lepton sector.
The flavor charge of the lepton doublet $k_{L_i}$ can be determined by the neutrino oscillation data for each hierarchical scheme. The difference of the flavor charge is defined as $\delta k_{Lij}\equiv k_{L_i}-k_{L_j}$. Then, the ratio between the mass of the active neutrinos is given by the flavor charge in the normal hierarchical spectrum:
\begin{align}
\epsilon^{2\delta k_{L23}} &= \frac{m_2}{m_3} \sim \dfrac{\sqrt{\Delta m_{21}^2}}{\sqrt{\Delta m_{21}^2+\Delta m_{32}^2}} \\
\therefore 2\delta k_{L23} &= \log_\epsilon \left( \dfrac{\sqrt{\Delta m_{21}^2}}{\sqrt{\Delta m_{21}^2+\Delta m_{32}^2}} \right) \sim 1.16.
\end{align}
Note that the prefactor $2$ comes from the fact that $m_{\mathrm{eff}}$ depends only on the flavor charge of the lepton doublet $k_{L_i}$.
We parametrize the difference $\delta k_{L23}=p=0,1$. Note that the mass of the lightest active neutrino cannot be determined by the experimental data, but it's implied by the ordering of the flavor charge $k_{L_1}\geq k_{L_2} \geq k_{L_3}$.

Then, according to the CKM mixing matrix, $\delta k_{L12}$ and $\delta k_{L13}$ are parametrized as following:
\begin{align}
&\delta k_{L12} = 1\\
&\delta k_{L13} = 1+p
\end{align}
By using the ansatz, we can evaluate the PMNS matrix in terms of the flavor charge,
\begin{align}
U_{\mathrm{PMNS}} \sim 
\begin{pmatrix}
1 & \epsilon^{1} & \epsilon^{1+p} \\
\epsilon^{1} & 1 & \epsilon^{p} \\
\epsilon^{1+p} & \epsilon^{p} & 1\\
\end{pmatrix}.
\end{align}
The ansatz of the flavor charge can be fixed by assigning the scale of the active neutrino mass for the largest component in the effective mass. The flavor charge should satisfy the following condition for each dominant contribution,
\begin{description}
\item[RHN] :
\begin{align}
\left( m_{\mathrm{eff}}\right)_{33} &= v\sin^2\beta\epsilon^{2+2(k_{H_u}+k_{L_3})+w-k_{N_3}} \sim 0.05 \mathrm{eV} \nonumber \\
\therefore k_{L_3} &\sim \dfrac{1}{2}\left(3k_{N_3} -2 -w + \log_\epsilon\left(\dfrac{m_3}{v}\right)\right) - k_{H_u} -k_{N_3}
\end{align}

\item[Neutralino] :
\begin{align}
\left( m_{\mathrm{eff}}\right)_{33} &= \dfrac{v}{1+\tan^2\beta}\epsilon^{2n_{33}^\nu-w} \sim 0.05 \mathrm{eV} \nonumber \\
\therefore k_{L_3} &\sim \dfrac{1}{2}\left(w + \log_\epsilon\left[ \left\{ 1+\left(\dfrac{m_t}{m_b}\right)^2\epsilon^{2x}\right\}\dfrac{m_3}{v}\right]\right) - k_{H_u} -k_{N_3}
\end{align}
\end{description}

Then, the charge of the lepton doublet can be reduced as 
\begin{align}
\begin{cases}
k_{L_1} = 1+p+k_{L_3}\\
k_{L_2} = p+k_{L_3}\\
k_{L_3} = l-k_{H_u}-k_{N_3},
\end{cases}
\end{align}
where $l$ parametrizes the scale of the SS contribution\footnote{In the analysis of the flavor charge, we take the approximated integer value of the parameter $l$.}, where 
\begin{align}
l=
\begin{cases}
\dfrac{1}{2}\left(3k_{N_3} -2 -w + \log_\epsilon\left(\dfrac{m_3}{v}\right)\right)\,\, ({\mathrm{RHN}})\\
\dfrac{1}{2}\left(w + \log_\epsilon\left[ \left\{ 1+\left(\dfrac{m_t}{m_b}\right)^2\epsilon^{2x}\right\}\dfrac{m_3}{v}\right]\right)\,\, ({\mathrm{neutralino}})
\end{cases}
\label{eq:lforNH}
\end{align}

For the inverted hierarchy, the ansatz of the flavor charge should be modified, 
\begin{align}
\begin{cases}
k_{L_1} = k_{L_2} = l-k_{H_u}-k_{N_3}\\
k_{L_3} = k_{L_2}+p,
\end{cases}
\label{eq:kLforIH}
\end{align}
where the parameter $l$ is defined as Eq.({\ref{eq:lforNH}}).

\section{Numerical Analysis}
\label{sec:Numerical Analysis}
\subsection{Anomaly cancellation and Flavor ansatz}
In this section, we numerically study the concrete assignment of the flavor symmetry. If we assume the ansatz for the Yukawa coupling and $\mu$-terms, the flavor charge is reduced into the parameters given in Tab.\ref{tb:DCtab1} for the normal hierarchy. For the inverted hierarchy, we use the ansatz of the lepton doublet $k_{L_i}$ following Eq.(\ref{eq:kLforIH}). In comparison with our previous work \cite{Hinata:2020cdt}, the flavor charge of the lepton doublet is reduced since those predict the matrix texture of the active neutrino.
\begin{table}[t]
\caption{The ansatz of the flavor charge (NH)}
\label{tb:charge constraint}
\centering
\begin{tabular}{rll}\hline\hline
&$k_{H_u}= -w - k_{H_d}$ & \\
&$k_{Q_2}=k_{Q_1}-1-y$ & \\
&$k_{Q_3}=k_{Q_1}-3-y$ & \\
&$k_{U_1}=-k_{H_u}-k_{Q_1}+8$ & \\
&$k_{U_2}=k_{U_1}-3+y$ & \\
&$k_{U_3}=k_{U_1}-5+y$ & \\
&$k_{D_1}=-k_{H_d}-k_{Q_1}+4+x$ & \\
&$k_{D_2}=k_{D_1}-1+y$ & \\
&$k_{D_3}=k_{D_1}-1+y$ & \\
&$k_{L_1} = 1+p+k_{L_3}$ &\\
&$k_{L_2} = 1+k_{L_3}$&\\
&$k_{L_3} = l-k_{H_u}-k_{N_3} $& \\
&$k_{E_1}=-k_{H_d}-k_{L_1}+4+x+z$ & \\
&$k_{E_2}=-k_{H_d}-k_{L_2}+2+x$ & \\
&$k_{E_3}=-k_{H_d}-k_{L_3}+x$ & \\
\hline\hline
\end{tabular}
\label{tb:DCtab1}
\end{table}

This is because the flavor charge that realizes the fermion mass hierarchy and mixing are determined by some parameters. In Tab.\ref{tb:DCtab2}, the parameters in the model and those corresponding quantities are shown. The flavor charge is also restricted to be validity of the seesaw approximation $n_{33}^\nu -2 w -4 \geq 0$ and to obtain the correct mass spectrum $-3 \leq 3-2x-2w+3k_{N_3}\leq 0$ in the neutralino dominated case.

Furthermore, the flavor charge is also constrained by the anomaly cancellation condition because of gauging the symmetry. We require that the mixed anomaly with the SM gauge groups has to cancel via the Green-Schwarz mechanism \cite{Green:1984sg}, where the anomaly coefficients are canceled by the shift of the single dilaton field. The condition to vanish the anomaly is given by the anomaly coefficients ${\cal A}_{\cdots}$ which are defined in Appendix.\ref{sec:anomaly},
\begin{align}
\dfrac{{\cal{A}}_{CCX}}{k_C}=\dfrac{{\cal{A}}_{WWX}}{k_W}=\dfrac{{\cal{A}}_{YYX}}{k_Y},
\label{eq:ACGS}
\end{align}
where the $k_{C,W,Y}$ are the Ka\v{c}-Moody level, and it decides the normalisation to embed the unified gauge group. It is an integer for the non-Abelian group, while it is allowed in rational number for Abelian group. If we consider the standard unification scenario, such as $\mathrm{SU}(5)$, the Ka\v{c}-Moody levels have to satisfy the following relation,
\begin{equation}
k_{C}=k_W=3k_Y/5=1.
\end{equation}
However, the non-canonical hypercharge normalization is required in the context of the gauge coupling unification, or derived from some orbifold models \cite{Dienes:1996du, PerezLorenzana:1999tf, Barger:2005qy, Gogoladze:2007ck}. Thus, we treat the hypercharge normalization $k_Y$ as parameter in the analysis of the anomaly cancellation, while we fix $k_W=k_C=1$. We restrict the parameter space in the rational number and $1\leq k_Y \leq3$.
\begin{table}[t]
\caption{List of parameters}
\label{tb:parameters}
\centering
\begin{tabular}{ccc}\hline\hline
parameter & range & corresponding quantities \\ \hline
$x$ & $0\sim 3$ & $\epsilon^x=\frac{m_b}{m_t}\tan\beta$ \\
$y$ & $0,1$ & the CKM matrix \\
$z$ & $0,1$ & charged lepton mass \\
$w$ & $1\sim 6$ & $\epsilon^{w} = v/m_{\mathrm{soft}}$ \\
$p$ & $0,1$ & the PMNS matrix \\
$k_Y$ & $1\leq k_Y \leq3$ & hypercharge normalization \\
$k_{N_{3}} $ & $1\sim 5$ & flavor charge of RHNs \\
$k_{H_{u\alpha}}, k_{H_{d\alpha}}$ & $-15\sim15$ & flavor charge of Ex. Higgs \\
$k_{H_{u\alpha}}+k_{H_{d\alpha}}=\Omega_\alpha/3 $ & $1\sim 4$ & mass scale of Ex. Higgs \\
\hline\hline
\end{tabular}
\label{tb:DCtab2}
\end{table}

In this parameter space, we search for the solution satisfying the anomaly cancellation condition Eq.(\ref{eq:ACGS}) and the flavor ansatz given in Tab.\ref{tb:DCtab1}. The solutions are divided into four cases: RHN dominated case, and neutralino dominated case for NH or IH, depending on the dominant contribution to $m_{\mathrm{eff}}$ and the flavor charge. We find the solutions to satisfy the anomaly cancellation conditions, and provide that the anomaly-free charge assignment in Appendix.\ref{sec:Charge}.

Interestingly, the flavor charge of the third generation of RHNs $k_{N_3}$ participates in the anomaly cancellation since it determines the mass scale of the Majorana mass term even if it is the SM singlet. On the other hand, the flavor charge of the first and second generation of RHN $k_{N_{1,2}}$ cannot be determined, thus it remains as the free parameter, which controls the mass scale of the light species of the sterile neutrino. 
The flavor charge of the extra-Higgs field is not uniquely determined by the above analysis, so we choose one example giving the same physical observables. 

We have to comment on the solution about the hypercharge normalization. For the given solution, there is no solution with the standard hypercharge normalization $k_Y=5/3$. Then, it is difficult to occur the gauge coupling unification of the SM gauge group, even if the extra-Higgs fields contribute to the running of the gauge coupling.

\subsection{Observables in Lepton Sector}
We analyze the Yukawa texture for the flavor charge which satisfies the anomaly cancellation condition. In addition to the SM sector, the mass spectrum of the RHNs can be calculated by fixing those flavor charges. For the 1st and 2nd generation of RHNs, the flavor charges are irrelevant to the anomaly cancellation and flavor observables, thus, we have to fix $k_{N_{1,2}}$ to discuss the model prediction. In this calculation, we fix $k_{N_1}=k_{N_3}+2$ and $k_{N_2}=k_{N_3}+1$.

Then, we evaluate the observables in the lepton sector, following the one example of the flavor charge in Tab.\ref{tb:fcharge eg}, where the charge assignment implies the RHN dominated for NH (No.1 in Appendix.\ref{sec:Charge}). We numerically calculate the flavor observables and fit ${\cal O}(1)$ coefficients of the Yukawa coupling.
\begin{table}
\caption{Flavor charge of the fields contents and parameters of the model. The charge assignment is derive by several parameters Eq.(\ref{eq:fcharge para eg}) which solve the anomaly cancellation conditions.}
\small
\centering
\begin{tabular}{cccccccccc}\hline\hline
 $X_{Q_{1,2,3}}$& $X_{U_{1,2,3}}$ & $X_{D_{1,2,3}}$ & $X_{L_{1,2,3}}$ & $X_{E_{1,2,3}}$ & $X_{N_{1,2,3}}$ & $X_{H_u}$ & $X_{H_d}$ & $X_{\Theta_{1}}$ \\ \hline
 $(6, 3, -3)$ & $(2,-7, -13)$ & $(31, 28, 28)$ & $(13,10,10)$ & $(24, 21, 15)$ & $(7,4,1)$ & $16$ & $-25$ & $-3$ \\ \hline\hline
\end{tabular}
\label{tb:fcharge eg}
\end{table}
In addition to the MSSM field, we have four pair of the extra-Higgs doublet fields and those flavor charges are determined by the anomaly cancellation,
\begin{align}
&X_{H_{u2,3,4,5}}=(-43,-34,-31,-16)\\
&X_{H_{d2,3,4,5}}=(46,46,43,28).
\end{align}
To suppress the mixing between the lepton doublet and $H_{ua}$, the flavor charge of $H_{ua}$ should be negative. 

In this example, the flavor charge is characterized by the following parameter set:
\begin{equation}
(k_{Q_1},k_{H_d},k_{N_3},x,y,z,w,p,\Omega,\Xi,k_Y)=(2, -8, 1, 0, 0, 0, 3, 0, 39, -2643, 65/54).
\label{eq:fcharge para eg}
\end{equation}
The parameters $\Omega$ and $\Xi$ are the contribution to the anomaly coefficients from the extra-Higgs doublets, which are defined in Appendix.\ref{sec:anomaly}.

In order to explain the observed mass and mixing angle in the lepton sector, we fit the ${\cal O}(1)$ factor in the superpotential Eq.(\ref{eq:WinfM3}). 
We fix the coefficients and numerically calculate the mass and mixing angle, then we carry out the $\chi^2$-analysis to evaluate the difference to the observed value $x^{\mathrm{obs}}_i$,
\begin{equation}
\chi^2 = \sum_i \left( \dfrac{x_i - x^\mathrm{obs}_i}{\sigma_i} \right)^2
\end{equation}
where $\sigma_i$ is the $1\sigma$ deviation while $x_i$ is the observables numerically evaluated. We can find the best fit values of the ${\cal O}(1)$ coefficients to minimize $\chi^2 \sim 0.36$ for the Yukawa matrix in lepton sector and $L$ violating operator,  
\begin{align}
y_e&=\begin{pmatrix}
 -1.12877 \epsilon ^4 & 0.791339 \epsilon ^3 & -1.27388 \epsilon ^3 \\
 1.48345 \epsilon ^3 & -1.89571 \epsilon ^2 & -0.494829 \epsilon ^2 \\
 0.28783 \epsilon  & -0.601413 & -0.59204 \\
\end{pmatrix},\\
y_\nu&=\begin{pmatrix}
 -1.1592 \epsilon ^{12} & 1.23835 \epsilon ^{11} & -0.86121 \epsilon^{11} \\
 -0.897902 \epsilon ^{11} & 0.702075 \epsilon ^{10} & 1.04136\epsilon^{10} \\
 1.98412 \epsilon ^{10} & -0.132766 \epsilon ^9 & -1.4796 \epsilon^9\\
\end{pmatrix},\\
\kappa_{ij1}&=
\begin{pmatrix}
-1.56861 \epsilon ^7 & 0.780551 \epsilon ^6 & 0.94564 \epsilon ^5 \\
 0.780551 \epsilon ^6 & -0.13229 \epsilon ^5 & 1.5355 \epsilon ^4 \\
 0.94564 \epsilon ^5 & 1.5355 \epsilon ^4 & 0.718014 \epsilon ^3 \\
\end{pmatrix}_{ij},\\
\kappa_{ij2}&=
\begin{pmatrix}
 0.780551 \epsilon ^6 & -0.13229 \epsilon ^5 & 1.5355 \epsilon ^4 \\
 -0.13229 \epsilon ^5 & -1.36367 \epsilon ^4 & 0.740113 \epsilon ^3 \\
 1.5355 \epsilon ^4 & 0.740113 \epsilon ^3 & 1.31348 \epsilon ^2 \\
\end{pmatrix}_{ij}, \\
\kappa_{ij3}&=
\begin{pmatrix}
 0.94564 \epsilon ^5 & 1.5355 \epsilon ^4 & 0.718014 \epsilon ^3 \\
 1.5355 \epsilon ^4 & 0.740113 \epsilon ^3 & 1.31348 \epsilon ^2 \\
 0.718014 \epsilon ^3 & 1.31348 \epsilon ^2 & -1.0699 \epsilon  \\
\end{pmatrix}_{ij}
\end{align}
Note that the dominant contribution of the effective Majorana mass matrix is the third component of $\kappa$ because of the smallest charge of RHNs.
The absolute value of maximum and minimum in the ${\cal O}(1)$ coefficients are $(r_{\mathrm{max}},r_{\mathrm{min}})=(1.984, 0.1323)$.

The VEVs of the Higgs doublet and the sneutrinos are also determined to realize the observables.
\begin{align}
&(v_d, v_u, v_{R_1},v_{R_2},v_{R_3} )=(246.1, 5.955, 2.669\times10^4, 2.716\times 10^4, 2.540\times10^4) \mathrm{GeV},\\
&(\phi_d, \phi_u, \phi_{R_1}, \phi_{R_2}, \phi_{R_3}) = ( -0.58, -0.26, -0.48, 0.38, 0.88)\pi.
\end{align}

\begingroup
\renewcommand{\arraystretch}{1.4}
\begin{table}
\caption{We show the neutrino oscillation parameter within the $1\sigma$ range for normal hierarchical spectrum \cite{Zyla:2020zbs,Esteban:2020cvm} and we provide the corresponding result of the numerical calculation by fixing the ${\cal O}(1)$ coefficients in the Yukawa matrices. The 4th and 5th column show the result of the RHN dominated case (No.1) and the neutralino dominated case (No.66), respectively.}
\centering
\begin{tabular}{|c|c|c||wc{30mm}|wc{30mm}|}\hline\hline
& obs. & \rm{bfp}$\pm 1\sigma$ & RHN(No.1)  & Neutralino(No.66)\\ \hline
%\multirow{2}{*}{mass ratio}& $m_e/m_\tau$ & $0.0002872 \pm 0.$ & 0.0002879 & 0.00028755\\ 
%& $m_\mu/m_\tau$ & $0.05948 \pm 0. $ & 0.05948 & 0.05948  \\ \hline\hline
\multirow{2}{*}{mass square[$\mathrm{eV}^2$]} & $\Delta m_{21}^2 \, \times10^{-5}$ & $7.42^{+0.21}_{-0.20}$ & $7.420$ & $7.418$ \\ 
& $\Delta m_{31}^2 \, \times10^{-3}$ & $2.515^{+0.028}_{-0.028}$ & $2.514$ & $2.514$ \\ \hline\hline
\multirow{3}{*}{leptonic mixing} & $\sin^2\theta_{12}$ & $0.304^{+0.013}_{-0.012}$ & $0.3040$ &  $0.3039$  \\ 
& $\sin^2\theta_{23}$ & $0.573^{+0.018}_{-0.023}$ & 0.5700 & 0.5699 \\
& $\sin^2\theta_{13}$ & $0.0222^{+0.00068}_{-0.00062}$ & 0.02221 & 0.02221 \\ \hline
\multirow{1}{*}{CP phase} & $\delta_{\mathrm{PMNS}}$ & $1.078^{+0.2889}_{-0.1389}$ & $1.00$ & $0.9437$ \\ \hline\hline
 & {$\chi^2$} &  - & 0.36 & 0.94 \\ \hline\hline
\end{tabular}
\label{tb:obsNH}
\end{table}
\endgroup

In Tab.\ref{tb:obsNH}, we show the model prediction of the neutrino oscillation parameters for a  charge assignment by using the given coupling constants. If we choose the flavor charge of the neutralino dominated case, we can also obtain the similar result of the neutrino oscillation parameters.
The mass of RHN $m_{si}$ is also obtained,
\begin{align}
(m_{s1},m_{s2},m_{s3}) = (12.20, 567.5, 6631.)\, \mathrm{GeV}.
\end{align}
In this assignment of the flavor charge, we can obtain the sterile neutrino below the $\mathrm{GeV}$ scale by choosing the flavor charge $k_{N_1}$. We give the phenomenological implication of the lightest sterile neutrino, in the next subsection, which may contribute to the neutrinoless double beta decay, or play a role of the dark matter in cosmology.

On the other hands, we show the numerical fitting result for the inverted hierarchy in Table.\ref{tb:obsIH}. Compared with the normal hierarchy case, $\chi^2$ is large because the CP phase is smaller than the center value. Note that the charged lepton mass hierarchy can be also realized in the given charge assignment, which does not depends on the hierarchical scheme or the dominated contribution.

Finally, we show the Higgs sector with the extra-Higgs boson. The supersymmetric Higgs mass matrix is originated from the two contribution: superpotential and K\"ahler potential.
\begin{align}\mu_{ij} \sim
\begin{pmatrix}
 \epsilon ^3 m_{\mathrm{soft}}  & 0 & 0 & 0 & 0 \\
 0 & \epsilon M_{\mathrm{Pl}}  & \epsilon M_{\mathrm{Pl}} & M_{\mathrm{Pl}} &  \epsilon^5m_{\mathrm{soft}}  \\
 0 & \epsilon^4M_{\mathrm{Pl}} & \epsilon^4M_{\mathrm{Pl}} & \epsilon^3M_{\mathrm{Pl}} &  \epsilon^2m_{\mathrm{soft}} \\
 0 & \epsilon^5M_{\mathrm{Pl}} & \epsilon^5M_{\mathrm{Pl}} & \epsilon^4M_{\mathrm{Pl}} &  \epsilon m_{\mathrm{soft}}  \\
 0 & \epsilon ^{10}M_{\mathrm{Pl}} & \epsilon ^{10}M_{\mathrm{Pl}} & \epsilon^9M_{\mathrm{Pl}} & \epsilon^4 M_{\mathrm{Pl}}  \\
\end{pmatrix}_{ij}
\end{align}
The extra-Higgs fields are enough heavy due to the flavor charge assignment, then those masses around $\sim 10^{15}{\mathrm{GeV}}$.

\begingroup
\renewcommand{\arraystretch}{1.4}
\begin{table}
\caption{We show the neutrino oscillation parameter within the $1\sigma$ range for inverted hierarchical spectrum \cite{Zyla:2020zbs,Esteban:2020cvm} and we provide the corresponding result of the numerical calculation by fixing the ${\cal O}(1)$ coefficients in the Yukawa matrices. The 4th and 5th column show the result of the RHN dominated case (No.80) and the neutralino dominated case (No.164), respectively.}
\centering
\begin{tabular}{|c|c|c||wc{30mm}|wc{30mm}|}\hline\hline
& obs. & \rm{bfp}$\pm 1\sigma$ & RHN(No.80)  & Neutralino(No.164)\\ \hline
%\multirow{2}{*}{mass ratio}& $m_e/m_\tau$ & $0.0002872 \pm 0.$ & 0.0002879 & 0.00028755\\ 
%& $m_\mu/m_\tau$ & $0.05948 \pm 0. $ & 0.05948 & 0.05948  \\ \hline\hline
\multirow{2}{*}{mass square[$\mathrm{eV}^2$]} & $\Delta m_{21}^2 \, \times10^{-5}$ & $7.42^{+0.21}_{-0.20}$ & $7.420$ & $7.421$ \\ 
& $\Delta m_{32}^2 \, \times10^{-3}$ & $-2.498^{+0.028}_{-0.029}$ & $-2.497$ & $-2.497$ \\ \hline\hline
\multirow{3}{*}{leptonic mixing} & $\sin^2\theta_{12}$ & $0.304^{+0.012}_{-0.012}$ & $0.3039$ &  $0.3033$  \\ 
& $\sin^2\theta_{23}$ & $0.578^{+0.017}_{-0.021}$ & 0.5749 & 0.5756 \\
& $\sin^2\theta_{13}$ & $0.02238^{+0.00064}_{-0.00062}$ & 0.02240 & 0.02240 \\ \hline
\multirow{1}{*}{CP phase} & $\delta_{\mathrm{PMNS}}$ & $1.594^{+0.15}_{-0.18}$ & $1.00$ & $1.00$ \\ \hline\hline
 & {$\chi^2$} &  - & 15.42 & 15.46\\ \hline\hline
\end{tabular}
\label{tb:obsIH}
\end{table}
\endgroup

\subsection{Phenomenological implication}
In our model, the flavor charges of the first and the second generation of the right-handed neutrino $k_{N_{1,2}}$ are not determined by the anomaly cancellation condition, and then, we can find that the active neutrino masses and mixing angles can be explained independently of the mass of the sterile neutrino. 

Depending on the choice of the flavor charge, the light sterile neutrino is predicted in our model, which can be applied to the other phenomenological issue in particle physics. Because of the Majorana mass of RHNs, it contributes to the neutrinoless double beta decay based on the lepton number violation term. On the other hands, if the lightest sterile neutrino has the mass around ${\mathrm{keV}}$, it may play an important role in cosmological observation as a candidate for dark matter (DM). 
In this section, we show that the lightest sterile neutrino appear depending on the flavor charges $k_{N_{1,2}}$, and discuss the phenomenological consequence of the choice of the flavor charge. We can find that in the $\mathrm{keV}$-scale sterile neutrino dark matter scenario, which is produced via the Dodelson-Widrow mechanism, we cannot explain the DM abundance. Furthermore, we show that some flavor charge assignments derived in the previous section can be excluded by the constraint on the active-sterile mixing and the non-observation of the neutrinoless double beta decay, if the sterile neutrino has the mass below few hundreds $\mathrm{MeV}$.

\subsubsection{$\mathrm{keV}$-scale sterile neutrino dark matter}
First of all, let us consider that the sterile neutrino plays a role of the dark matter. In the recent studies, the sterile neutrino can be regarded as a candidate for the warm dark matter \cite{Dodelson:1993je, Shi:1998km, Abazajian:2001nj, Shaposhnikov:2006xi, Asaka:2006rw, Asaka:2006nq, Abada:2014zra, Boyarsky:2018tvu, Abazajian:2021zui}. In our model, the $\mathrm{keV}$-scale sterile neutrino can be predicted by choosing the flavor charge.

The DM sterile neutrino is produced through the mixing between the active and the sterile neutrino. If the primordial asymmetry of the total lepton number is small, the relic abundance of the DM sterile neutrino can be evaluated by the mass of the sterile neutrino $m_{s1}$ and the mixing angle $\Theta=\theta^\ast U_{N}$ \cite{Dodelson:1993je}. In our model, the effective coupling constants are provided by the flavor charge, therefore, we can evaluate the mass and the mixing angle in terms of the flavor charge. In order to compare with the DM abundance, we adopt the approximated form of the production of the sterile neutrino \cite{Abazajian:2001nj},
\begin{equation}
   \Omega_{N} h^{2}=0.3\left(\frac{\sin ^{2} 2 \theta}{10^{-10}}\right)\left(\frac{m_{s1}}{100 \mathrm{keV}}\right)^{2},
\end{equation}
where the active-sterile mixing angle is defined by $\sin^2 2\theta= \sum_{i=e,\mu,\tau}|\Theta_{i 1}|^2$. 

The mixing angle $\Theta_{i 1}$ is defined by the Dirac mass and the Majorana mass of RHN, and we can express in terms of the flavor charge,
\begin{align}
\Theta_{i 1} &= \theta_{i j}^\ast\left(U_{N}\right)_{j 1} \sim \epsilon^{2+w-k_{N_{3}}+k_{H u}+k_{L_{i}}-k_{N_{1}}}\leq \epsilon^{2+w-k_{N_{3}}+k_{H u}+k_{L_{3}}-k_{N_{1}}},
\end{align}
where we neglect the unitary matrix diagonalizing the Majorana mass matrix. The last inequality follows from the normal hierarchical scheme $k_{L_1}\leq k_{L_2}\leq k_{L_3}$. In our analysis, the squared-mass difference of the active neutrino can be realized by the two seesaw contribution, then the flavor charge of the lepton doublet $k_{L_i}$ depends on the parameter $l$ which determines the mass scale of the active neutrino (see also Eq.(\ref{eq:lforNH})).
On the other hands, the mass of the lightest species of the sterile neutrino is evaluated by the Majorana mass term,
\begin{equation}
    m_{s1}\sim\left(M_{R}\right)_{11}=v \epsilon^{-2-w+k_{N_{3}}+2 k_{N_{1}}} \leq 100\mathrm{keV}.
\end{equation}

By using these expression, we find the relic abundance $\Omega_Nh^2$ in terms of the flavor charge only depends on the power of the neutrino Yukawa coupling $y^\nu_{ij}\sim \epsilon^{n_{13}^\nu}$,
\begin{equation}
    \Omega_N h^2 \sim 0.3\times 10^{22} \times \epsilon^{2n_{13}^\nu}.
    \label{eq:RA}
\end{equation}

In order to explain the dark matter abundance by the $\mathrm{keV}$-scale sterile neutrino, the relic abundance (\ref{eq:RA}) have to be suppressed to avoid the overproduction. 
\begin{align}
    m_{s1}\leq 100\mathrm{keV}&\rightarrow -(2+w-k_{N_3})+2k_{N_1}\geq 10 \\
    \sum_i|\Theta_{i1}|^2\leq 10^{-10}&\rightarrow  (2+w-k_{N_3})+k_{H_u}+k_{L_i}-k_{N_1}\geq 8 
\end{align}

Combining with the solution of the anomaly-free flavor charge given in Appendix.\ref{sec:Charge}, there is no solution to explain simultaneously the DM abundance and the neutrino observables. 
Intuitively, we can find that $m_{s1}$ and $\Theta_{i1}$ have the opposite dependence of $k_{N_1}$. For the RHN dominated case, the constraints of $m_{s1}$ and $\Theta_{i1}$ are not simultaneously valid. In the case of the neutralino dominated case, while there is room for the parameter that satisfy the constrains of them to hold at the same time, such choice of the parameter is excluded by the anomaly cancellation conditions and the flavor observables. Therefore, we can conclude that the $\mathrm{keV}$-scale sterile neutrino cannot realize the relic abundance by the Dodelson-Widrow scenario.

\subsubsection{Neutrinoless double beta decay}
Another possibility to detect the nature of the Majorana neutrino is the neutrinoless double beta ($0\nu\beta\beta$) decay which is one of the most fascinating signal of the lepton number violation. Since the $0\nu\beta\beta$ decay is caused by the existence of the Majorana mass term of the neutrino, the observation of the $0\nu\beta\beta$ decay provides the severe constraints on the flavor charge in our model. 

\begin{figure}[t]
 \centering
\includegraphics[width=14.0cm]{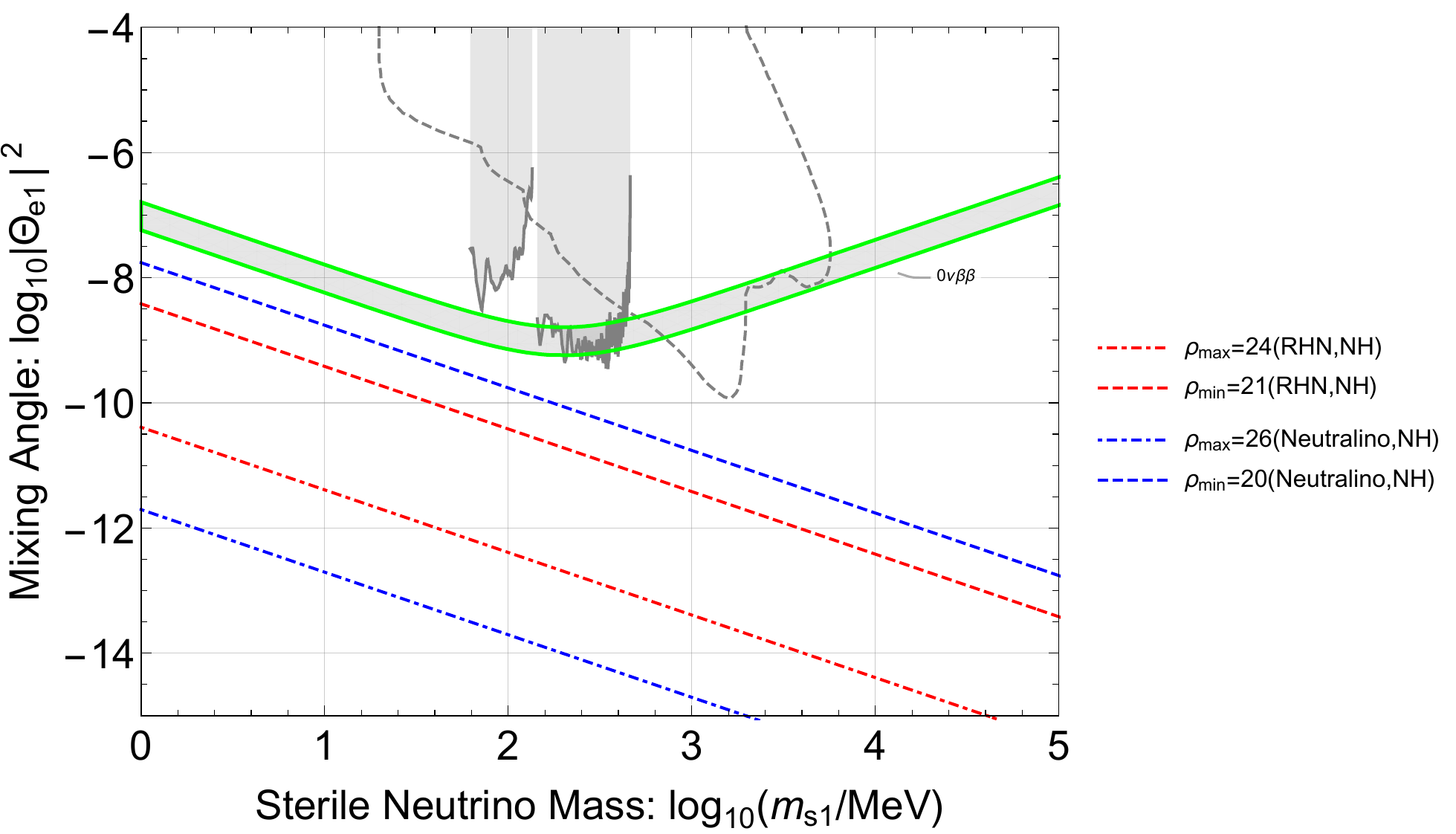}
  \caption{The sterile neutrino mass $m_{s1}$ and the active-sterile mixing $\Theta_{e1}$ given the anomaly free flavor charge in normal hierarchical scheme. The red plots show the RHN contribution dominates the effective active neutrino, while the blue ones show the neutralino contribution dominates it. The dashed (dot-dashed) means the choice of the flavor charge maximize (minimize) the power of ($m_{s1}|\Theta_{e1}|^2$). The Green band shows the constraints from the $0\nu\beta\beta$ decay. The gray shadows (dashed line) are the current (future) sensitivity bounds of $\Theta_{e1}$ given in heavy neutral lepton searches \cite{PIENU:2011aa,PIENU:2017wbj,NA62:2020mcv,SHiP:2018xqw}.}
  \label{fig:NH}
\end{figure}

\begin{figure}[t]
 \centering
\includegraphics[width=14.0cm]{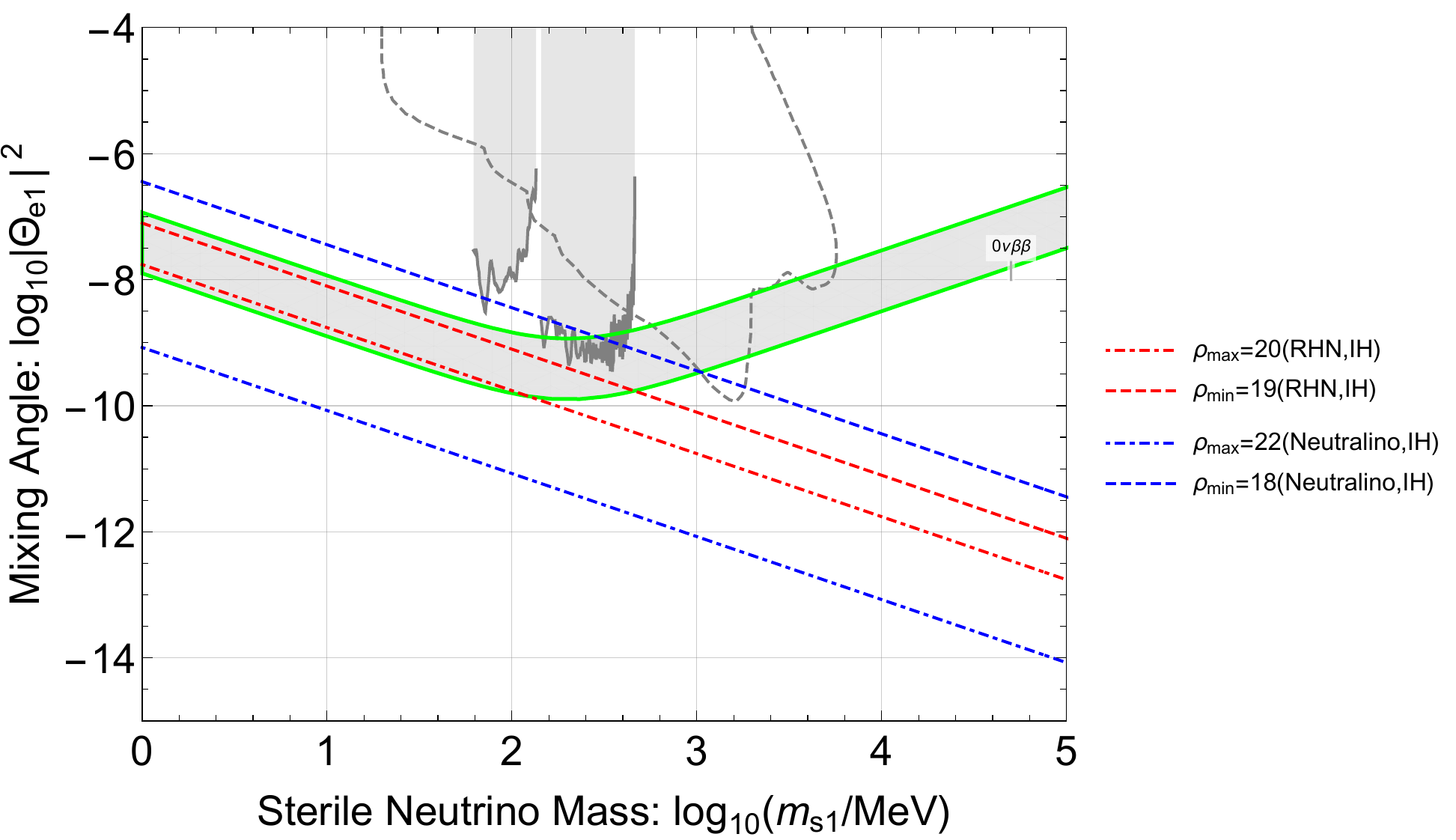}
  \caption{The sterile neutrino mass $m_{s1}$ and the active-sterile mixing $\Theta_{e1}$ given the anomaly free flavor charge in inverted hierarchical scheme. The red plots show the RHN contribution dominates the effective active neutrino, while the blue ones show the neutralino contribution dominates it. The dashed (dot-dashed) means the choice of the flavor charge maximize (minimize) the power of ($m_{s1}|\Theta_{e1}|^2$). The Green band shows the constraints from the $0\nu\beta\beta$ decay. The gray shadows (dashed line) are the current (future) sensitivity bounds of $\Theta_{e1}$ given in heavy neutral lepton searches \cite{PIENU:2011aa,PIENU:2017wbj,NA62:2020mcv,SHiP:2018xqw}.}
  \label{fig:IH}
\end{figure}
The $0\nu\beta\beta$ have not been observed and the half-life of the decay is severely constrained by the KamLAND-ZEN with $^{136}\mathrm{Xe}$, $\tau_{1/2}>1.07\times 10^{26}\mathrm{yr}$ at $90\%$ CL \cite{KamLAND-Zen:2016pfg}. 
The existence of the Majorana mass term at the low energy induces the $0\nu\beta\beta$ decay, then the half lifetime of the $0\nu\beta\beta$ decay $\tau_{1/2}$ is given by 
\begin{equation}
    \tau_{1/2}^{-1} = G|{\cal M}m_{ee}|,
\end{equation}
where $G$ is the phase space integral and ${\cal M}$ is the nuclear matrix element of the active neutrino. The effective Majorana mass $m_{ee}$ can be divided into the contribution from the active and sterile neutrino,
\begin{align}
    m_{ee}&=m_{ee}^\nu+m_{ee}^N \nonumber \\
    &=\sum_i (U_{\mathrm{PMNS}})_{ei}^2 m_{i} + \sum_i \Theta_{ei}^2 m_{si}f_{\beta}(m_{si}),
    \label{eq:mee}
\end{align}
where the function $f_{\beta}(m_{si})$ represents the suppression factor by the propagator of the RH-neutrino \cite{Faessler:2014kka,Barea:2015zfa}. It can be expressed by using the Fermi momentum $\Lambda_\beta$ depending on the nucleus of each experiments,
\begin{equation}
 f_{\beta}(m_{si}) = \dfrac{\Lambda_\beta^2}{m_{si}^2+\Lambda_\beta^2}.
\end{equation}
In the following analysis, we set $\Lambda_\beta=200\mathrm{MeV}$ as a reference value. In general, the additional contribution to the $0\nu\beta\beta$ decay is induced in the supersymmetric SM, or in RpV scenario \cite{Mohapatra:1986su,Hirsch:1995vr,Hirsch:1997dm}. In our model, the RpV terms except $\kappa$ are prohibited by $M_3$, and it is known that the contribution by the propagation of the sparticle is suppressed, comparing with the one of the Majorana mass. The lepton number violating coupling $\kappa$ also affects on the $0\nu\beta\beta$ decay at the loop level, but such diagram is suppressed by the loop factor. Thus, we can estimate the effective Majorana mass which is given by Eq.(\ref{eq:mee}).

By using a variety of nuclear matrix element calculations, we can estimate the upper bound on the effective Majorana mass,
\begin{equation}
    m_{ee}<61-165\mathrm{meV}.
\end{equation}
Furthermore, combining with the squared-mass difference and the mixing angle of the active neutrino, we find that the contribution from the effective Majorana mass of active neutrino $m_{ee}^\nu$ can be estimated as $|m_{ee}^\nu|=1.45-3.68 \mathrm{meV}$ and $18.6-48.4\mathrm{meV}$ for the NH and IH cases, respectively. 

Again, we consider the second and third generation of the sterile neutrino is heavy enough to decouple at the low energy effective theory. By using the approximate form of $m_{s1}$ and $\Theta_{e1}$, the solution of the flavor charge predict the effective Majorana mass term for the RH-neutrino. While $m_{s1}$ and $\Theta_{e1}$ depend on $k_{N_i}$, the dependence is cancelled in the effective mass except the suppression factor $f_{\beta}(m_{si})$. The effective mass is suppressed by $f_{\beta}(m_{si})$ when the mass of RHN is large, and the maximum contribution to the effective Majorana mass is given when $k_{N_i}$ is large.

In Fig.\ref{fig:NH} and Fig.\ref{fig:IH}, we plot the sterile neutrino mass $m_{s1}$ and the active-sterile mixing $\Theta_{e1}$ given the anomaly free flavor charge given in Appendix.\ref{sec:Charge} for each hierarchical scheme. For each plot, the red plots show the RHN contribution dominates the effective active neutrino, while the blue ones show the neutralino contribution dominates it. The dashed (dot-dashed) means the choice of the flavor charge maximize (minimize) the power of ($m_{s1}|\Theta_{e1}|^2$), which is parametrized by $\rho$ defined by
 \begin{equation}
     \rho \equiv 2+w-k_{N_3}+2(k_{H_u}+k_{L_1}) \quad  \left(m_{s1}|\Theta_{e1}|^2 \sim v\epsilon^\rho\right),
 \end{equation}
 where the product $m_{s1}$ and $|\Theta_{e1}|^2$ is independent of $k_{N_1}$, and the prediction of the mixing angle varies with the given $m_{s1}$ for each model.
 The green band shows the constraints from the $0\nu\beta\beta$ decay when the contribution from the active neutrino $m_{ee}^\nu$ is maximized. For the upper line corresponds to the conservative bounds for $m_{ee} < 165\mathrm{meV}$, while the other corresponds to $m_{ee}<61\mathrm{meV}$. The gray shadows (dashed line) are the current (future) sensitivity bounds of $\Theta_{e1}$ given in heavy neutral lepton searches \cite{PIENU:2011aa,PIENU:2017wbj,NA62:2020mcv,SHiP:2018xqw}.

For normal hierarchical scheme, the mixing angle is suppressed and the sterile neutrino mass is not restricted by the observation. On the other hands, for inverted hierarchical scheme, the flavor charge $k_{N_1}$, which predicts the lightest mass of sterile neutrino, is restricted the $0\nu\beta\beta$ decay and the search for the heavy neutral lepton, even if the RHN is irrelevant to the anomaly cancellation. There are some assignments to avoid the observational constraint in the neutralino dominated case compared with the RHN one since the neutralino dominated case tends to be large SUSY breaking scale $m_{\mathrm{soft}}\sim v\epsilon^{-w}$, and to be large Majorana mass scale $v\epsilon^{2+w-k_{N_3}}$ to suppress the RHN contribution. Note that we have to include the ${\cal O}(1)$ coefficients of the coupling in the superpotential in order to realize the neutrino mass and mixing angle.

\section{Conclusion}
\label{sec:Conclusion}
The matter triality is one of the Abelian discrete symmetries that suppresses the proton decay up to the dimension-5 operator in the supersymmetric Lagrangian even if the R-parity is violated. This symmetry naturally requires the three generations of the right-handed neutrinos in the viewpoint of the discrete anomaly cancellation. In this paper, we analyze the Type-I-like seesaw mechanism in SSM based on the matter triality. The lepton number violation term $\kappa_{ijk}$ appears under $M_3$, and it provides the Majorana mass via the VEV of the RH-sneutrino. The effective mass of the active neutrino is generated from the contribution of the Majorana mass, and the neutralino participates in the seesaw mechanism due to the R-parity violation. 

We construct the flavor model based on the Froggatt-Nielsen mechanism to explain the fermion mass hierarchy and the mixing. To realize the correct mass hierarchy and the mixing, the flavor charges of the matter fields are strictly constrained. In addition to the Yukawa texture, the gauge anomaly cancellation conditions are assigned to embed the flavor symmetry into the gauge symmetry $\mathrm{U}(1)_X$. 

The flavor symmetry can control the mass spectrum of the neutrino for the NH and IH. The active neutrino mass matrix can be evaluated in terms of the flavor charge, and we find there exist the two contributions to $m_{\mathrm{eff}}$, {\it i.e.} RHN or the neutralino, depending on the flavor charge. We derive the solutions which are consistent with the anomaly cancellation condition, and we make sure that the solution certainly realizes the mass differences, the leptonic mixing angles, and the CP phase. Then, we show that the sterile neutrino mass and the mixing of the sterile neutrino with the active neutrino depends on the flavor charge. Especially, for inverted hierarchical scheme, the flavor charge $k_{N_1}$, which predicts the lightest mass of sterile neutrino, is restricted the $0\nu\beta\beta$ decay and the search for the heavy neutral lepton, even if the RHN is irrelevant to the anomaly cancellation. On the other hands, the mixing angle $\Theta_{e1}$ is suppressed in some of the neutralino dominated case compared with the RHN one since the neutralino dominated case tends to be large SUSY breaking scale $m_{\mathrm{soft}}\sim v\epsilon^{-w}$, and to be large Majorana mass scale $v\epsilon^{2+w-k_{N_3}}$ to suppress the RHN contribution.
As a result, the seesaw mechanism is enlarged in the SSM with $M_3$ because of the existence of the neutralino dominated case.  
Although the most of the flavor charge assignments avoid the heavy neutral lepton search above $\mathrm{MeV}$-scale, we find the $\mathrm{keV}$-scale sterile neutrino DM cannot be realized in the Dodelson-Widrow scenario.

In addition to the existence of the additional states of the neutrino, the RpV signal is also of great interest. The RpV term $\kappa_{ijk}$ should be constrained by the wash-out of the baryon number such as the other RpV terms \cite{Fischler:1990gn, Dreiner:1992vm}. We would like to explore the detectability in comparison with the other supersymmetric model with/without R-parity in future work. 

%--------section6-------------%

\section*{Acknowledgments}
We would like to thank H. Abe for helpful discussions and advice.

%--------Appendix------------%
\appendix
\section{The Mixing Angle in the Seesaw Mechanism}
\label{sec:mixing}
The mixing angle $\theta$ and $\varepsilon$ can be evaluated in terms of the flavor charge.
\begin{align}
\theta_{ij} =&\left(m_D^T M_R^{-1}\right)_{ij} \sim \epsilon^{2+w+k_{H_u}-k_{N_3}-k_{N_j}+k_{L_i}}, \label{eq:theta_ana} \\
\varepsilon_{ij} =& \left( -M_N^{-1}M_X\right) 
\sim \dfrac{\left(\sum_l \epsilon^{2k_{N_l}} \right)}{\mu \left\{c_\beta s_\beta(B^2M_1+A^2M_2)v^2+M_1M_2\mu \right\} }c_i\epsilon^{k_{H_u}+k_{L_j}-w},
\end{align}
where $n_{ij}^\nu =k_{H_u}+k_{N_j}+k_{L_i}$ and the coefficients $c_i$ depends on the input of the model, 
\begin{align}
c_i=
\begin{pmatrix} 
-Ac_\beta M_2v^2\mu, & Bc_\beta M_1v^2\mu, & -v(c_\beta s_\beta(B^2M_1+A^2M_2)v^2+M_1M_2\mu), & -c_\beta^2(B^2M_1+A^2M_2)v^3 
\end{pmatrix}.
\end{align}This analytic expression shows that the opposite dependence of the soft SUSY breaking mass $m_{\mathrm{soft}}\sim v\epsilon^{-w}$. 
Similarly, the mixing angles between RHN and neutralino are also obtained in terms of the flavr charge,
\begin{align}
& \eta_{ij} =\left(\varepsilon\theta^\ast\right)_{ij} \sim\dfrac{\left(\sum_l \epsilon^{2k_{L_l}} \right)}{\mu \left\{c_\beta s_\beta(B^2M_1+A^2M_2)v^2+M_1M_2\mu \right\} }c_i\epsilon^{2+k_{H_u}-k_{N_i}}, \\
&\xi_{i4}= \left( M_R^{-1}\theta^\dagger M_X^T\right)_{i4} \sim \epsilon^{4+2k_{H_u}-k_{N_i}-k_{N_3}+w}\left(\sum_l \epsilon^{2k_{L_l}} \right)\left(\sum_l \epsilon^{-2k_{N_l}} \right), \\
&\xi_{ia}= \left( M_R^{-1}\theta^\dagger M_X^T\right)_{ia}=0 \,\, (a=1,2,3) .
\end{align}

\section{Anomaly Coefficients}
\label{sec:anomaly}
The anomaly coefficients are evaluated to combine with the flavor charge by using the Fujikawa method \cite{Fujikawa:1979ay}. To solve the anomaly cancellation, we introduce the pair of the extra-Higgs doublets. 
The representation of the Higgs doublets is given under the gauge group $\mathrm{SU}(2)_L \times \mathrm{U}(1)_Y \times \mathrm{U}(1)_X$,
\begin{align}
H_{ua} :& \, \, \left({\bf 2}, \,\, +\frac{1}{2}, \,\, 2+3k_{H_{ua}} \right) \\
H_{da} :& \, \, \left({\bf 2}, \,\, -\frac{1}{2}, \,\, 1+3k_{H_{da}} \right),
\end{align}
where $k_{H_{ua}}$ and $k_{H_{da}} $ are the flavor charge of the up- or down-type extra-Higgs doublet.
Then, the anomaly coefficients can be evaluated in terms of the flavor charge,
\begin{align}
& {\cal A}_{CCX} = \sum_i[{2X_{Qi} +X_{Ui}+X_{Di}}],\nonumber \\
& {\cal A}_{WWX} = \sum_i[{3X_{Qi} +X_{Li}}]+X_{H_u}+X_{H_d}+\Omega,\nonumber \\
& {\cal A}_{YYX} = \frac{1}{6}\sum_i[X_{Qi}+8X_{Ui}+2X_{Di}+3X_{Li}+6X_{Ei}] +\frac{1}{2}(X_{H_u}+X_{H_d})+\Omega,\nonumber \\
&{\cal A}_{YXX} = \sum_i[X_{Qi}^2-2X_{Ui}^2+X_{Di}^2-X_{Li}^2+X_{Ei}^2] -X_{H_u}^2+X_{H_d}^2))+\Xi.\nonumber 
\end{align}
The anomaly ${\cal A}_{CCX} $, ${\cal A}_{WWX} $, and ${\cal A}_{YYX} $ can be canceled by the Green-Schwarz mechanism. On the other hand, ${\cal A}_{YXX} $ cannot be canceled by the shift of the dilaton field, thus we require that 
\begin{equation}
{\cal A}_{YXX} =0.
\end{equation}
The contribution from the extra-Higgs fields can be evaluated as the parameters,
\begin{align}
\Omega=\sum_{a=2}^{N_h}\Omega_a,\quad \Omega_a/3 = {k_{H_{ua}}} + {k_{H_{ua}}} = 1\sim 4.
\end{align}
In addition to the scale of the extra-Higgs fields, the hypercharge anomaly ${\cal{A}}_{YXX}$ include the contribution,
\begin{align}
\Xi=\sum_{a=2}^{N_h}\Xi_a, \quad \Xi_a= 3\Omega_a(-2+3({k_{H_{ua}}}-{k_{H_{da}}})). 
\end{align}

\section{The Flavor Charge Assignment}
\label{sec:Charge}
 We provide the solution to the anomaly cancellation conditions and the flavor structure for the quark and lepton. If we choose the set of parameters, then we can derive the concrete flavor charge. The parameters space and the physical implication are written in Tab.\ref{tb:DCtab2}. The Tab.\ref{tb:fcharge1}, \ref{tb:fcharge2}, \ref{tb:fcharge3} and \ref{tb:fcharge4} show the solution of the anomaly free flavor charge. The column "SS" means the dominant contribution in the effective active neutrino mass matrix $m_{\mathrm{eff}}$ which can be determined by the ratio $R$. Furthermore, we also show the parameters $\rho$, which characterize the relationship between $m_{s1}$ and $\Theta_{e1}$.
 The parameter $\rho$ is defined by 
 \begin{equation}
     \rho \equiv 2+w-k_{N_3}+2(k_{H_u}+k_{L_1}) \quad  \left(m_{s1}|\Theta_{e1}|^2 \sim v\epsilon^\rho\right),
 \end{equation}
 where the product $m_{s1}$ and $|\Theta_{e1}|^2$ is independent of $k_{N_1}$, and the prediction of the mixing angle varies with the given $m_{s1}$ for each model.
\begin{table}[htb]
\caption{The solution of the anomaly free flavor charge}
\label{tb:fcharge1}
\footnotesize
\centering
\begin{tabular}{ccccccccccccccccc}\hline\hline
No.	& SS & Hierarchy & $k_{Q_1}$ & $k_{H_d}$ & $k_{N_3}$ & $x$ & $y$ & $z$ & $w$ & $p$ & $\Omega$ & $\Xi$ & $k_Y$ &   $\rho$ & $0\nu\beta\beta$ \\ \hline	
1	&	 RHN	&	 NO	& $	2	$ & $	-8	$ & $	1	$ & $	0	$ & $	0	$ & $	0	$ & $	3	$ & $	0	$ & $	39	$ & $	-2643	$ & $	65	/	54	$ & $	22	$ & $	\sqrt{}	$ \\
2	&	 RHN	&	 NO	& $	2	$ & $	-8	$ & $	1	$ & $	0	$ & $	0	$ & $	0	$ & $	3	$ & $	1	$ & $	33	$ & $	-2235	$ & $	61	/	54	$ & $	24	$ & $	\sqrt{}	$ \\
3	&	 RHN	&	 NO	& $	2	$ & $	-7	$ & $	1	$ & $	0	$ & $	0	$ & $	1	$ & $	3	$ & $	0	$ & $	30	$ & $	-1644	$ & $	61	/	54	$ & $	22	$ & $	\sqrt{}	$ \\
4	&	 RHN	&	 NO	& $	2	$ & $	-7	$ & $	1	$ & $	0	$ & $	0	$ & $	1	$ & $	3	$ & $	1	$ & $	24	$ & $	-1254	$ & $	19	/	18	$ & $	24	$ & $	\sqrt{}	$ \\
5	&	 RHN	&	 NO	& $	2	$ & $	-8	$ & $	2	$ & $	0	$ & $	0	$ & $	0	$ & $	3	$ & $	0	$ & $	39	$ & $	-2643	$ & $	65	/	54	$ & $	21	$ & $	\sqrt{}	$ \\
6	&	 RHN	&	 NO	& $	2	$ & $	-8	$ & $	2	$ & $	0	$ & $	0	$ & $	0	$ & $	3	$ & $	1	$ & $	33	$ & $	-2235	$ & $	61	/	54	$ & $	23	$ & $	\sqrt{}	$ \\
7	&	 RHN	&	 NO	& $	2	$ & $	-7	$ & $	2	$ & $	0	$ & $	0	$ & $	1	$ & $	3	$ & $	0	$ & $	30	$ & $	-1644	$ & $	61	/	54	$ & $	21	$ & $	\sqrt{}	$ \\
8	&	 RHN	&	 NO	& $	2	$ & $	-7	$ & $	2	$ & $	0	$ & $	0	$ & $	1	$ & $	3	$ & $	1	$ & $	24	$ & $	-1254	$ & $	19	/	18	$ & $	23	$ & $	\sqrt{}	$ \\
9	&	 RHN	&	 NO	& $	2	$ & $	-7	$ & $	2	$ & $	1	$ & $	0	$ & $	0	$ & $	3	$ & $	0	$ & $	39	$ & $	-2301	$ & $	7	/	6	$ & $	21	$ & $	\sqrt{}	$ \\
10	&	 RHN	&	 NO	& $	2	$ & $	-7	$ & $	2	$ & $	1	$ & $	0	$ & $	0	$ & $	3	$ & $	1	$ & $	33	$ & $	-1893	$ & $	11	/	10	$ & $	23	$ & $	\sqrt{}	$ \\
11	&	 RHN	&	 NO	& $	2	$ & $	-8	$ & $	2	$ & $	1	$ & $	0	$ & $	1	$ & $	3	$ & $	1	$ & $	42	$ & $	-3288	$ & $	37	/	30	$ & $	23	$ & $	\sqrt{}	$ \\
12	&	 RHN	&	 NO	& $	3	$ & $	-7	$ & $	2	$ & $	1	$ & $	1	$ & $	1	$ & $	3	$ & $	0	$ & $	30	$ & $	-2346	$ & $	11	/	10	$ & $	21	$ & $	\sqrt{}	$ \\
13	&	 RHN	&	 NO	& $	3	$ & $	-7	$ & $	2	$ & $	1	$ & $	1	$ & $	1	$ & $	3	$ & $	1	$ & $	24	$ & $	-1920	$ & $	31	/	30	$ & $	23	$ & $	\sqrt{}	$ \\
14	&	 RHN	&	 NO	& $	3	$ & $	-10	$ & $	3	$ & $	0	$ & $	0	$ & $	0	$ & $	6	$ & $	1	$ & $	42	$ & $	-3180	$ & $	19	/	18	$ & $	23	$ & $	\sqrt{}	$ \\
15	&	 RHN	&	 NO	& $	2	$ & $	-8	$ & $	3	$ & $	0	$ & $	0	$ & $	1	$ & $	3	$ & $	0	$ & $	30	$ & $	-2220	$ & $	61	/	54	$ & $	22	$ & $	\sqrt{}	$ \\
16	&	 RHN	&	 NO	& $	2	$ & $	-8	$ & $	3	$ & $	0	$ & $	0	$ & $	1	$ & $	3	$ & $	1	$ & $	24	$ & $	-1794	$ & $	19	/	18	$ & $	24	$ & $	\sqrt{}	$ \\
17	&	 RHN	&	 NO	& $	3	$ & $	-9	$ & $	3	$ & $	0	$ & $	0	$ & $	1	$ & $	6	$ & $	0	$ & $	39	$ & $	-2481	$ & $	19	/	18	$ & $	21	$ & $	\sqrt{}	$ \\
18	&	 RHN	&	 NO	& $	3	$ & $	-9	$ & $	3	$ & $	0	$ & $	0	$ & $	1	$ & $	6	$ & $	1	$ & $	33	$ & $	-2019	$ & $	1			$ & $	23	$ & $	\sqrt{}	$ \\
19	&	 RHN	&	 NO	& $	2	$ & $	-8	$ & $	3	$ & $	0	$ & $	1	$ & $	1	$ & $	3	$ & $	1	$ & $	42	$ & $	-2370	$ & $	23	/	18	$ & $	24	$ & $	\sqrt{}	$ \\
20	&	 RHN	&	 NO	& $	2	$ & $	-8	$ & $	3	$ & $	1	$ & $	0	$ & $	0	$ & $	3	$ & $	0	$ & $	39	$ & $	-2967	$ & $	7	/	6	$ & $	22	$ & $	\sqrt{}	$ \\
21	&	 RHN	&	 NO	& $	2	$ & $	-8	$ & $	3	$ & $	1	$ & $	0	$ & $	0	$ & $	3	$ & $	1	$ & $	33	$ & $	-2523	$ & $	11	/	10	$ & $	24	$ & $	\sqrt{}	$ \\
22	&	 RHN	&	 NO	& $	2	$ & $	-7	$ & $	3	$ & $	1	$ & $	0	$ & $	1	$ & $	3	$ & $	0	$ & $	30	$ & $	-1860	$ & $	11	/	10	$ & $	22	$ & $	\sqrt{}	$ \\
23	&	 RHN	&	 NO	& $	2	$ & $	-7	$ & $	3	$ & $	1	$ & $	0	$ & $	1	$ & $	3	$ & $	1	$ & $	24	$ & $	-1434	$ & $	31	/	30	$ & $	24	$ & $	\sqrt{}	$ \\
24	&	 RHN	&	 NO	& $	2	$ & $	-7	$ & $	3	$ & $	2	$ & $	0	$ & $	0	$ & $	3	$ & $	0	$ & $	39	$ & $	-2571	$ & $	25	/	22	$ & $	22	$ & $	\sqrt{}	$ \\
25	&	 RHN	&	 NO	& $	2	$ & $	-7	$ & $	3	$ & $	2	$ & $	0	$ & $	0	$ & $	3	$ & $	1	$ & $	33	$ & $	-2127	$ & $	71	/	66	$ & $	24	$ & $	\sqrt{}	$ \\
26	&	 RHN	&	 NO	& $	2	$ & $	-7	$ & $	3	$ & $	3	$ & $	0	$ & $	1	$ & $	3	$ & $	1	$ & $	42	$ & $	-3162	$ & $	7	/	6	$ & $	24	$ & $	\sqrt{}	$ \\
27	&	 RHN	&	 NO	& $	3	$ & $	-6	$ & $	3	$ & $	3	$ & $	1	$ & $	1	$ & $	3	$ & $	0	$ & $	30	$ & $	-2112	$ & $	19	/	18	$ & $	22	$ & $	\sqrt{}	$ \\
28	&	 RHN	&	 NO	& $	3	$ & $	-6	$ & $	3	$ & $	3	$ & $	1	$ & $	1	$ & $	3	$ & $	1	$ & $	24	$ & $	-1650	$ & $	1			$ & $	24	$ & $	\sqrt{}	$ \\
29	&	 RHN	&	 NO	& $	2	$ & $	-8	$ & $	4	$ & $	0	$ & $	0	$ & $	1	$ & $	3	$ & $	0	$ & $	30	$ & $	-2220	$ & $	61	/	54	$ & $	21	$ & $	\sqrt{}	$ \\
30	&	 RHN	&	 NO	& $	2	$ & $	-8	$ & $	4	$ & $	0	$ & $	0	$ & $	1	$ & $	3	$ & $	1	$ & $	24	$ & $	-1794	$ & $	19	/	18	$ & $	23	$ & $	\sqrt{}	$ \\
31	&	 RHN	&	 NO	& $	3	$ & $	-10	$ & $	4	$ & $	0	$ & $	0	$ & $	1	$ & $	6	$ & $	1	$ & $	33	$ & $	-2667	$ & $	1			$ & $	24	$ & $	\sqrt{}	$ \\
32	&	 RHN	&	 NO	& $	2	$ & $	-8	$ & $	4	$ & $	0	$ & $	1	$ & $	1	$ & $	3	$ & $	1	$ & $	42	$ & $	-2370	$ & $	23	/	18	$ & $	23	$ & $	\sqrt{}	$ \\
33	&	 RHN	&	 NO	& $	2	$ & $	-8	$ & $	4	$ & $	1	$ & $	0	$ & $	0	$ & $	3	$ & $	0	$ & $	39	$ & $	-2967	$ & $	7	/	6	$ & $	21	$ & $	\sqrt{}	$ \\
34	&	 RHN	&	 NO	& $	2	$ & $	-8	$ & $	4	$ & $	1	$ & $	0	$ & $	0	$ & $	3	$ & $	1	$ & $	33	$ & $	-2523	$ & $	11	/	10	$ & $	23	$ & $	\sqrt{}	$ \\
35	&	 RHN	&	 NO	& $	2	$ & $	-7	$ & $	4	$ & $	1	$ & $	0	$ & $	1	$ & $	3	$ & $	0	$ & $	30	$ & $	-1860	$ & $	11	/	10	$ & $	21	$ & $	\sqrt{}	$ \\
36	&	 RHN	&	 NO	& $	2	$ & $	-7	$ & $	4	$ & $	1	$ & $	0	$ & $	1	$ & $	3	$ & $	1	$ & $	24	$ & $	-1434	$ & $	31	/	30	$ & $	23	$ & $	\sqrt{}	$ \\
37	&	 RHN	&	 NO	& $	3	$ & $	-9	$ & $	4	$ & $	1	$ & $	0	$ & $	1	$ & $	6	$ & $	0	$ & $	39	$ & $	-2751	$ & $	27	/	26	$ & $	22	$ & $	\sqrt{}	$ \\
38	&	 RHN	&	 NO	& $	2	$ & $	-7	$ & $	4	$ & $	2	$ & $	0	$ & $	0	$ & $	3	$ & $	0	$ & $	39	$ & $	-2571	$ & $	25	/	22	$ & $	21	$ & $	\sqrt{}	$ \\
39	&	 RHN	&	 NO	& $	2	$ & $	-7	$ & $	4	$ & $	2	$ & $	0	$ & $	0	$ & $	3	$ & $	1	$ & $	33	$ & $	-2127	$ & $	71	/	66	$ & $	23	$ & $	\sqrt{}	$ \\
40	&	 RHN	&	 NO	& $	2	$ & $	-7	$ & $	4	$ & $	3	$ & $	0	$ & $	1	$ & $	3	$ & $	1	$ & $	42	$ & $	-3162	$ & $	7	/	6	$ & $	23	$ & $	\sqrt{}	$ \\
41	&	 RHN	&	 NO	& $	3	$ & $	-6	$ & $	4	$ & $	3	$ & $	1	$ & $	1	$ & $	3	$ & $	0	$ & $	30	$ & $	-2112	$ & $	19	/	18	$ & $	21	$ & $	\sqrt{}	$ \\
42	&	 RHN	&	 NO	& $	3	$ & $	-6	$ & $	4	$ & $	3	$ & $	1	$ & $	1	$ & $	3	$ & $	1	$ & $	24	$ & $	-1650	$ & $	1			$ & $	23	$ & $	\sqrt{}	$ \\
43	&	 RHN	&	 NO	& $	3	$ & $	-10	$ & $	5	$ & $	0	$ & $	0	$ & $	1	$ & $	6	$ & $	1	$ & $	33	$ & $	-2667	$ & $	1			$ & $	23	$ & $	\sqrt{}	$ \\
44	&	 RHN	&	 NO	& $	2	$ & $	-8	$ & $	5	$ & $	0	$ & $	1	$ & $	0	$ & $	3	$ & $	0	$ & $	39	$ & $	-2103	$ & $	65	/	54	$ & $	22	$ & $	\sqrt{}	$ \\
45	&	 RHN	&	 NO	& $	2	$ & $	-9	$ & $	5	$ & $	0	$ & $	1	$ & $	1	$ & $	3	$ & $	0	$ & $	48	$ & $	-3480	$ & $	73	/	54	$ & $	22	$ & $	\sqrt{}	$ \\
46	&	 RHN	&	 NO	& $	2	$ & $	-9	$ & $	5	$ & $	0	$ & $	1	$ & $	1	$ & $	3	$ & $	1	$ & $	42	$ & $	-3018	$ & $	23	/	18	$ & $	24	$ & $	\sqrt{}	$ \\
47	&	 RHN	&	 NO	& $	2	$ & $	-8	$ & $	5	$ & $	1	$ & $	0	$ & $	1	$ & $	3	$ & $	0	$ & $	30	$ & $	-2490	$ & $	11	/	10	$ & $	22	$ & $	\sqrt{}	$ \\
48	&	 RHN	&	 NO	& $	2	$ & $	-8	$ & $	5	$ & $	1	$ & $	0	$ & $	1	$ & $	3	$ & $	1	$ & $	24	$ & $	-2028	$ & $	31	/	30	$ & $	24	$ & $	\sqrt{}	$ \\
49	&	 RHN	&	 NO	& $	3	$ & $	-9	$ & $	5	$ & $	1	$ & $	0	$ & $	1	$ & $	6	$ & $	0	$ & $	39	$ & $	-2751	$ & $	27	/	26	$ & $	21	$ & $	\sqrt{}	$ \\
50	&	 RHN	&	 NO	& $	2	$ & $	-8	$ & $	5	$ & $	1	$ & $	1	$ & $	1	$ & $	3	$ & $	1	$ & $	42	$ & $	-2640	$ & $	37	/	30	$ & $	24	$ & $	\sqrt{}	$ \\ \hline\hline
\end{tabular}
\end{table}

\begin{table}[htb]
\caption{The solution of the anomaly free flavor charge (conti.)}
\label{tb:fcharge2}
\footnotesize
\centering
\begin{tabular}{ccccccccccccccccc}\hline\hline
No.	& SS & Hierarchy & $k_{Q_1}$ & $k_{H_d}$ & $k_{N_3}$ & $x$ & $y$ & $z$ & $w$ & $p$ & $\Omega$ & $\Xi$ & $k_Y$ &   $\rho$ & $0\nu\beta\beta$ \\ \hline	
51	&	 RHN	&	 NO	& $	3	$ & $	-9	$ & $	5	$ & $	2	$ & $	0	$ & $	0	$ & $	6	$ & $	1	$ & $	42	$ & $	-3072	$ & $	43	/	42	$ & $	23	$ & $	\sqrt{}	$ \\	
52	&	 RHN	&	 NO	& $	2	$ & $	-7	$ & $	5	$ & $	2	$ & $	0	$ & $	1	$ & $	3	$ & $	0	$ & $	30	$ & $	-2076	$ & $	71	/	66	$ & $	22	$ & $	\sqrt{}	$ \\	
53	&	 RHN	&	 NO	& $	2	$ & $	-7	$ & $	5	$ & $	2	$ & $	0	$ & $	1	$ & $	3	$ & $	1	$ & $	24	$ & $	-1614	$ & $	67	/	66	$ & $	24	$ & $	\sqrt{}	$ \\	
54	&	 RHN	&	 NO	& $	2	$ & $	-7	$ & $	5	$ & $	3	$ & $	0	$ & $	0	$ & $	3	$ & $	0	$ & $	39	$ & $	-2841	$ & $	10	/	9	$ & $	22	$ & $	\sqrt{}	$ \\	
55	&	 RHN	&	 NO	& $	2	$ & $	-7	$ & $	5	$ & $	3	$ & $	0	$ & $	0	$ & $	3	$ & $	1	$ & $	33	$ & $	-2361	$ & $	19	/	18	$ & $	24	$ & $	\sqrt{}	$ \\	
56	&	 RHN	&	 NO	& $	4	$ & $	-8	$ & $	5	$ & $	3	$ & $	1	$ & $	1	$ & $	6	$ & $	0	$ & $	39	$ & $	-3111	$ & $	91	/	90	$ & $	21	$ & $	\sqrt{}	$ \\	
57	&	 Neutralino	&	 NO	& $	2	$ & $	-8	$ & $	1	$ & $	0	$ & $	0	$ & $	0	$ & $	3	$ & $	0	$ & $	39	$ & $	-2643	$ & $	65	/	54	$ & $	22	$ & $	\sqrt{}	$ \\	
58	&	 Neutralino	&	 NO	& $	2	$ & $	-8	$ & $	1	$ & $	0	$ & $	0	$ & $	0	$ & $	3	$ & $	1	$ & $	33	$ & $	-2235	$ & $	61	/	54	$ & $	24	$ & $	\sqrt{}	$ \\	
59	&	 Neutralino	&	 NO	& $	2	$ & $	-7	$ & $	1	$ & $	0	$ & $	0	$ & $	1	$ & $	3	$ & $	0	$ & $	30	$ & $	-1644	$ & $	61	/	54	$ & $	22	$ & $	\sqrt{}	$ \\	
60	&	 Neutralino	&	 NO	& $	2	$ & $	-7	$ & $	1	$ & $	0	$ & $	0	$ & $	1	$ & $	3	$ & $	1	$ & $	24	$ & $	-1254	$ & $	19	/	18	$ & $	24	$ & $	\sqrt{}	$ \\	
61	&	 Neutralino	&	 NO	& $	2	$ & $	-8	$ & $	1	$ & $	1	$ & $	0	$ & $	0	$ & $	3	$ & $	0	$ & $	39	$ & $	-2967	$ & $	7	/	6	$ & $	24	$ & $	\sqrt{}	$ \\	
62	&	 Neutralino	&	 NO	& $	2	$ & $	-8	$ & $	1	$ & $	1	$ & $	0	$ & $	0	$ & $	3	$ & $	1	$ & $	33	$ & $	-2523	$ & $	11	/	10	$ & $	26	$ & $	\sqrt{}	$ \\	
63	&	 Neutralino	&	 NO	& $	2	$ & $	-7	$ & $	1	$ & $	1	$ & $	0	$ & $	1	$ & $	3	$ & $	0	$ & $	30	$ & $	-1860	$ & $	11	/	10	$ & $	24	$ & $	\sqrt{}	$ \\	
64	&	 Neutralino	&	 NO	& $	2	$ & $	-7	$ & $	1	$ & $	1	$ & $	0	$ & $	1	$ & $	3	$ & $	1	$ & $	24	$ & $	-1434	$ & $	31	/	30	$ & $	26	$ & $	\sqrt{}	$ \\	
65	&	 Neutralino	&	 NO	& $	3	$ & $	-10	$ & $	2	$ & $	0	$ & $	0	$ & $	1	$ & $	6	$ & $	1	$ & $	33	$ & $	-2667	$ & $	1			$ & $	26	$ & $	\sqrt{}	$ \\	
66	&	 Neutralino	&	 NO	& $	2	$ & $	-7	$ & $	2	$ & $	2	$ & $	0	$ & $	0	$ & $	3	$ & $	0	$ & $	39	$ & $	-2571	$ & $	25	/	22	$ & $	23	$ & $	\sqrt{}	$ \\	
67	&	 Neutralino	&	 NO	& $	2	$ & $	-7	$ & $	2	$ & $	2	$ & $	0	$ & $	0	$ & $	3	$ & $	1	$ & $	33	$ & $	-2127	$ & $	71	/	66	$ & $	25	$ & $	\sqrt{}	$ \\	
68	&	 Neutralino	&	 NO	& $	2	$ & $	-7	$ & $	2	$ & $	3	$ & $	0	$ & $	1	$ & $	3	$ & $	1	$ & $	42	$ & $	-3162	$ & $	7	/	6	$ & $	25	$ & $	\sqrt{}	$ \\	
69	&	 Neutralino	&	 NO	& $	3	$ & $	-6	$ & $	2	$ & $	3	$ & $	1	$ & $	1	$ & $	3	$ & $	0	$ & $	30	$ & $	-2112	$ & $	19	/	18	$ & $	23	$ & $	\sqrt{}	$ \\	
70	&	 Neutralino	&	 NO	& $	3	$ & $	-6	$ & $	2	$ & $	3	$ & $	1	$ & $	1	$ & $	3	$ & $	1	$ & $	24	$ & $	-1650	$ & $	1			$ & $	25	$ & $	\sqrt{}	$ \\	
71	&	 Neutralino	&	 NO	& $	3	$ & $	-10	$ & $	3	$ & $	0	$ & $	0	$ & $	0	$ & $	6	$ & $	1	$ & $	42	$ & $	-3180	$ & $	19	/	18	$ & $	23	$ & $	\sqrt{}	$ \\	
72	&	 Neutralino	&	 NO	& $	3	$ & $	-9	$ & $	3	$ & $	0	$ & $	0	$ & $	1	$ & $	6	$ & $	0	$ & $	39	$ & $	-2481	$ & $	19	/	18	$ & $	21	$ & $	\sqrt{}	$ \\	
73	&	 Neutralino	&	 NO	& $	3	$ & $	-9	$ & $	3	$ & $	0	$ & $	0	$ & $	1	$ & $	6	$ & $	1	$ & $	33	$ & $	-2019	$ & $	1			$ & $	23	$ & $	\sqrt{}	$ \\	
74	&	 Neutralino	&	 NO	& $	3	$ & $	-9	$ & $	3	$ & $	1	$ & $	0	$ & $	1	$ & $	6	$ & $	0	$ & $	39	$ & $	-2751	$ & $	27	/	26	$ & $	23	$ & $	\sqrt{}	$ \\	
75	&	 Neutralino	&	 NO	& $	3	$ & $	-6	$ & $	3	$ & $	3	$ & $	1	$ & $	0	$ & $	3	$ & $	0	$ & $	39	$ & $	-2607	$ & $	10	/	9	$ & $	20	$ & $	\sqrt{}	$ \\	
76	&	 Neutralino	&	 NO	& $	3	$ & $	-6	$ & $	3	$ & $	3	$ & $	1	$ & $	0	$ & $	3	$ & $	1	$ & $	33	$ & $	-2163	$ & $	19	/	18	$ & $	22	$ & $	\sqrt{}	$ \\	
77	&	 Neutralino	&	 NO	& $	3	$ & $	-9	$ & $	4	$ & $	2	$ & $	0	$ & $	0	$ & $	6	$ & $	1	$ & $	42	$ & $	-3072	$ & $	43	/	42	$ & $	24	$ & $	\sqrt{}	$ \\	
78	&	 Neutralino	&	 NO	& $	3	$ & $	-9	$ & $	4	$ & $	3	$ & $	0	$ & $	0	$ & $	6	$ & $	1	$ & $	42	$ & $	-3360	$ & $	91	/	90	$ & $	26	$ & $	\sqrt{}	$ \\	
79	&	 Neutralino	&	 NO	& $	4	$ & $	-8	$ & $	5	$ & $	3	$ & $	1	$ & $	1	$ & $	6	$ & $	0	$ & $	39	$ & $	-3111	$ & $	91	/	90	$ & $	21	$ & $	\sqrt{}	$ \\	
80	&	 RHN	&	 IO	& $	2	$ & $	-7	$ & $	1	$ & $	0	$ & $	0	$ & $	0	$ & $	3	$ & $	0	$ & $	33	$ & $	-1731	$ & $	61	/	54	$ & $	20	$ & $	\quad	$ \\	
81	&	 RHN	&	 IO	& $	2	$ & $	-8	$ & $	1	$ & $	0	$ & $	0	$ & $	0	$ & $	3	$ & $	1	$ & $	39	$ & $	-2715	$ & $	65	/	54	$ & $	20	$ & $	\quad	$ \\	
82	&	 RHN	&	 IO	& $	2	$ & $	-8	$ & $	1	$ & $	0	$ & $	0	$ & $	1	$ & $	3	$ & $	0	$ & $	42	$ & $	-3036	$ & $	23	/	18	$ & $	20	$ & $	\quad	$ \\	
83	&	 RHN	&	 IO	& $	2	$ & $	-7	$ & $	1	$ & $	0	$ & $	0	$ & $	1	$ & $	3	$ & $	1	$ & $	30	$ & $	-1734	$ & $	61	/	54	$ & $	20	$ & $	\quad	$ \\	
84	&	 RHN	&	 IO	& $	3	$ & $	-8	$ & $	1	$ & $	0	$ & $	1	$ & $	0	$ & $	3	$ & $	0	$ & $	33	$ & $	-2739	$ & $	61	/	54	$ & $	20	$ & $	\quad	$ \\	
85	&	 RHN	&	 IO	& $	3	$ & $	-7	$ & $	1	$ & $	0	$ & $	1	$ & $	1	$ & $	3	$ & $	0	$ & $	24	$ & $	-1794	$ & $	19	/	18	$ & $	20	$ & $	\quad	$ \\	
86	&	 RHN	&	 IO	& $	2	$ & $	-7	$ & $	2	$ & $	0	$ & $	0	$ & $	0	$ & $	3	$ & $	0	$ & $	33	$ & $	-1731	$ & $	61	/	54	$ & $	19	$ & $	\quad	$ \\	
87	&	 RHN	&	 IO	& $	2	$ & $	-8	$ & $	2	$ & $	0	$ & $	0	$ & $	0	$ & $	3	$ & $	1	$ & $	39	$ & $	-2715	$ & $	65	/	54	$ & $	19	$ & $	\quad	$ \\	
88	&	 RHN	&	 IO	& $	2	$ & $	-8	$ & $	2	$ & $	0	$ & $	0	$ & $	1	$ & $	3	$ & $	0	$ & $	42	$ & $	-3036	$ & $	23	/	18	$ & $	19	$ & $	\quad	$ \\	
89	&	 RHN	&	 IO	& $	2	$ & $	-7	$ & $	2	$ & $	0	$ & $	0	$ & $	1	$ & $	3	$ & $	1	$ & $	30	$ & $	-1734	$ & $	61	/	54	$ & $	19	$ & $	\quad	$ \\	
90	&	 RHN	&	 IO	& $	3	$ & $	-8	$ & $	2	$ & $	0	$ & $	1	$ & $	0	$ & $	3	$ & $	0	$ & $	33	$ & $	-2739	$ & $	61	/	54	$ & $	19	$ & $	\quad	$ \\	
91	&	 RHN	&	 IO	& $	3	$ & $	-7	$ & $	2	$ & $	0	$ & $	1	$ & $	1	$ & $	3	$ & $	0	$ & $	24	$ & $	-1794	$ & $	19	/	18	$ & $	19	$ & $	\quad	$ \\	
92	&	 RHN	&	 IO	& $	2	$ & $	-7	$ & $	2	$ & $	1	$ & $	0	$ & $	0	$ & $	3	$ & $	1	$ & $	39	$ & $	-2373	$ & $	7	/	6	$ & $	19	$ & $	\quad	$ \\	
93	&	 RHN	&	 IO	& $	2	$ & $	-7	$ & $	2	$ & $	1	$ & $	0	$ & $	1	$ & $	3	$ & $	0	$ & $	42	$ & $	-2676	$ & $	37	/	30	$ & $	19	$ & $	\quad	$ \\	
94	&	 RHN	&	 IO	& $	2	$ & $	-8	$ & $	2	$ & $	1	$ & $	0	$ & $	1	$ & $	3	$ & $	1	$ & $	48	$ & $	-3840	$ & $	13	/	10	$ & $	19	$ & $	\quad	$ \\	
95	&	 RHN	&	 IO	& $	3	$ & $	-7	$ & $	2	$ & $	1	$ & $	1	$ & $	0	$ & $	3	$ & $	0	$ & $	33	$ & $	-2433	$ & $	11	/	10	$ & $	19	$ & $	\quad	$ \\	
96	&	 RHN	&	 IO	& $	3	$ & $	-6	$ & $	2	$ & $	1	$ & $	1	$ & $	1	$ & $	3	$ & $	0	$ & $	24	$ & $	-1416	$ & $	31	/	30	$ & $	19	$ & $	\quad	$ \\	
97	&	 RHN	&	 IO	& $	3	$ & $	-7	$ & $	2	$ & $	1	$ & $	1	$ & $	1	$ & $	3	$ & $	1	$ & $	30	$ & $	-2436	$ & $	11	/	10	$ & $	19	$ & $	\quad	$ \\	
98	&	 RHN	&	 IO	& $	2	$ & $	-8	$ & $	3	$ & $	0	$ & $	0	$ & $	0	$ & $	3	$ & $	0	$ & $	33	$ & $	-2307	$ & $	61	/	54	$ & $	20	$ & $	\quad	$ \\	
99	&	 RHN	&	 IO	& $	3	$ & $	-9	$ & $	3	$ & $	0	$ & $	0	$ & $	0	$ & $	6	$ & $	0	$ & $	42	$ & $	-2568	$ & $	19	/	18	$ & $	19	$ & $	\quad	$ \\	
100	&	 RHN	&	 IO	& $	2	$ & $	-7	$ & $	3	$ & $	0	$ & $	0	$ & $	1	$ & $	3	$ & $	0	$ & $	24	$ & $	-1344	$ & $	19	/	18	$ & $	20	$ & $	\quad	$ \\	
\hline\hline
\end{tabular}
\end{table}

\begin{table}[htb]
\caption{The solution of the anomaly free flavor charge (conti.2)}
\label{tb:fcharge3}
\footnotesize
\centering
\begin{tabular}{ccccccccccccccccc}\hline\hline
No.	& SS & Hierarchy & $k_{Q_1}$ & $k_{H_d}$ & $k_{N_3}$ & $x$ & $y$ & $z$ & $w$ & $p$ & $\Omega$ & $\Xi$ & $k_Y$ &   $\rho$ & $0\nu\beta\beta$ \\ \hline	
101	&	 RHN	&	 IO	& $	2	$ & $	-8	$ & $	3	$ & $	0	$ & $	0	$ & $	1	$ & $	3	$ & $	1	$ & $	30	$ & $	-2310	$ & $	61	/	54	$ & $	20	$ & $	\quad	$ \\
102	&	 RHN	&	 IO	& $	3	$ & $	-9	$ & $	3	$ & $	0	$ & $	0	$ & $	1	$ & $	6	$ & $	1	$ & $	39	$ & $	-2571	$ & $	19	/	18	$ & $	19	$ & $	\quad	$ \\
103	&	 RHN	&	 IO	& $	4	$ & $	-9	$ & $	3	$ & $	0	$ & $	1	$ & $	1	$ & $	6	$ & $	0	$ & $	33	$ & $	-2667	$ & $	1			$ & $	19	$ & $	\quad	$ \\
104	&	 RHN	&	 IO	& $	2	$ & $	-7	$ & $	3	$ & $	1	$ & $	0	$ & $	0	$ & $	3	$ & $	0	$ & $	33	$ & $	-1965	$ & $	11	/	10	$ & $	20	$ & $	\quad	$ \\
105	&	 RHN	&	 IO	& $	2	$ & $	-8	$ & $	3	$ & $	1	$ & $	0	$ & $	0	$ & $	3	$ & $	1	$ & $	39	$ & $	-3039	$ & $	7	/	6	$ & $	20	$ & $	\quad	$ \\
106	&	 RHN	&	 IO	& $	2	$ & $	-7	$ & $	3	$ & $	1	$ & $	0	$ & $	1	$ & $	3	$ & $	1	$ & $	30	$ & $	-1950	$ & $	11	/	10	$ & $	20	$ & $	\quad	$ \\
107	&	 RHN	&	 IO	& $	3	$ & $	-7	$ & $	3	$ & $	1	$ & $	1	$ & $	1	$ & $	3	$ & $	0	$ & $	24	$ & $	-2010	$ & $	31	/	30	$ & $	20	$ & $	\quad	$ \\
108	&	 RHN	&	 IO	& $	2	$ & $	-7	$ & $	3	$ & $	2	$ & $	0	$ & $	0	$ & $	3	$ & $	1	$ & $	39	$ & $	-2643	$ & $	25	/	22	$ & $	20	$ & $	\quad	$ \\
109	&	 RHN	&	 IO	& $	2	$ & $	-7	$ & $	3	$ & $	2	$ & $	0	$ & $	1	$ & $	3	$ & $	0	$ & $	42	$ & $	-2964	$ & $	79	/	66	$ & $	20	$ & $	\quad	$ \\
110	&	 RHN	&	 IO	& $	3	$ & $	-7	$ & $	3	$ & $	2	$ & $	1	$ & $	0	$ & $	3	$ & $	0	$ & $	33	$ & $	-2703	$ & $	71	/	66	$ & $	20	$ & $	\quad	$ \\
111	&	 RHN	&	 IO	& $	3	$ & $	-6	$ & $	3	$ & $	2	$ & $	1	$ & $	1	$ & $	3	$ & $	0	$ & $	24	$ & $	-1578	$ & $	67	/	66	$ & $	20	$ & $	\quad	$ \\
112	&	 RHN	&	 IO	& $	2	$ & $	-6	$ & $	3	$ & $	3	$ & $	0	$ & $	1	$ & $	3	$ & $	0	$ & $	42	$ & $	-2442	$ & $	7	/	6	$ & $	20	$ & $	\quad	$ \\
113	&	 RHN	&	 IO	& $	3	$ & $	-6	$ & $	3	$ & $	3	$ & $	1	$ & $	0	$ & $	3	$ & $	0	$ & $	33	$ & $	-2235	$ & $	19	/	18	$ & $	20	$ & $	\quad	$ \\
114	&	 RHN	&	 IO	& $	3	$ & $	-6	$ & $	3	$ & $	3	$ & $	1	$ & $	1	$ & $	3	$ & $	1	$ & $	30	$ & $	-2202	$ & $	19	/	18	$ & $	20	$ & $	\quad	$ \\
115	&	 RHN	&	 IO	& $	2	$ & $	-8	$ & $	4	$ & $	0	$ & $	0	$ & $	0	$ & $	3	$ & $	0	$ & $	33	$ & $	-2307	$ & $	61	/	54	$ & $	19	$ & $	\quad	$ \\
116	&	 RHN	&	 IO	& $	3	$ & $	-10	$ & $	4	$ & $	0	$ & $	0	$ & $	0	$ & $	6	$ & $	0	$ & $	42	$ & $	-3252	$ & $	19	/	18	$ & $	20	$ & $	\quad	$ \\
117	&	 RHN	&	 IO	& $	2	$ & $	-7	$ & $	4	$ & $	0	$ & $	0	$ & $	1	$ & $	3	$ & $	0	$ & $	24	$ & $	-1344	$ & $	19	/	18	$ & $	19	$ & $	\quad	$ \\
118	&	 RHN	&	 IO	& $	2	$ & $	-8	$ & $	4	$ & $	0	$ & $	0	$ & $	1	$ & $	3	$ & $	1	$ & $	30	$ & $	-2310	$ & $	61	/	54	$ & $	19	$ & $	\quad	$ \\
119	&	 RHN	&	 IO	& $	3	$ & $	-9	$ & $	4	$ & $	0	$ & $	0	$ & $	1	$ & $	6	$ & $	0	$ & $	33	$ & $	-2109	$ & $	1			$ & $	20	$ & $	\quad	$ \\
120	&	 RHN	&	 IO	& $	2	$ & $	-7	$ & $	4	$ & $	1	$ & $	0	$ & $	0	$ & $	3	$ & $	0	$ & $	33	$ & $	-1965	$ & $	11	/	10	$ & $	19	$ & $	\quad	$ \\
121	&	 RHN	&	 IO	& $	2	$ & $	-8	$ & $	4	$ & $	1	$ & $	0	$ & $	0	$ & $	3	$ & $	1	$ & $	39	$ & $	-3039	$ & $	7	/	6	$ & $	19	$ & $	\quad	$ \\
122	&	 RHN	&	 IO	& $	3	$ & $	-9	$ & $	4	$ & $	1	$ & $	0	$ & $	0	$ & $	6	$ & $	0	$ & $	42	$ & $	-2856	$ & $	27	/	26	$ & $	20	$ & $	\quad	$ \\
123	&	 RHN	&	 IO	& $	2	$ & $	-7	$ & $	4	$ & $	1	$ & $	0	$ & $	1	$ & $	3	$ & $	1	$ & $	30	$ & $	-1950	$ & $	11	/	10	$ & $	19	$ & $	\quad	$ \\
124	&	 RHN	&	 IO	& $	3	$ & $	-9	$ & $	4	$ & $	1	$ & $	0	$ & $	1	$ & $	6	$ & $	1	$ & $	39	$ & $	-2841	$ & $	27	/	26	$ & $	20	$ & $	\quad	$ \\
125	&	 RHN	&	 IO	& $	3	$ & $	-7	$ & $	4	$ & $	1	$ & $	1	$ & $	1	$ & $	3	$ & $	0	$ & $	24	$ & $	-2010	$ & $	31	/	30	$ & $	19	$ & $	\quad	$ \\
126	&	 RHN	&	 IO	& $	2	$ & $	-7	$ & $	4	$ & $	2	$ & $	0	$ & $	0	$ & $	3	$ & $	1	$ & $	39	$ & $	-2643	$ & $	25	/	22	$ & $	19	$ & $	\quad	$ \\
127	&	 RHN	&	 IO	& $	2	$ & $	-7	$ & $	4	$ & $	2	$ & $	0	$ & $	1	$ & $	3	$ & $	0	$ & $	42	$ & $	-2964	$ & $	79	/	66	$ & $	19	$ & $	\quad	$ \\
128	&	 RHN	&	 IO	& $	3	$ & $	-7	$ & $	4	$ & $	2	$ & $	1	$ & $	0	$ & $	3	$ & $	0	$ & $	33	$ & $	-2703	$ & $	71	/	66	$ & $	19	$ & $	\quad	$ \\
129	&	 RHN	&	 IO	& $	3	$ & $	-6	$ & $	4	$ & $	2	$ & $	1	$ & $	1	$ & $	3	$ & $	0	$ & $	24	$ & $	-1578	$ & $	67	/	66	$ & $	19	$ & $	\quad	$ \\
130	&	 RHN	&	 IO	& $	2	$ & $	-6	$ & $	4	$ & $	3	$ & $	0	$ & $	1	$ & $	3	$ & $	0	$ & $	42	$ & $	-2442	$ & $	7	/	6	$ & $	19	$ & $	\quad	$ \\
131	&	 RHN	&	 IO	& $	3	$ & $	-6	$ & $	4	$ & $	3	$ & $	1	$ & $	0	$ & $	3	$ & $	0	$ & $	33	$ & $	-2235	$ & $	19	/	18	$ & $	19	$ & $	\quad	$ \\
132	&	 RHN	&	 IO	& $	3	$ & $	-6	$ & $	4	$ & $	3	$ & $	1	$ & $	1	$ & $	3	$ & $	1	$ & $	30	$ & $	-2202	$ & $	19	/	18	$ & $	19	$ & $	\quad	$ \\
133	&	 RHN	&	 IO	& $	3	$ & $	-10	$ & $	5	$ & $	0	$ & $	0	$ & $	0	$ & $	6	$ & $	0	$ & $	42	$ & $	-3252	$ & $	19	/	18	$ & $	19	$ & $	\quad	$ \\
134	&	 RHN	&	 IO	& $	2	$ & $	-8	$ & $	5	$ & $	0	$ & $	0	$ & $	1	$ & $	3	$ & $	0	$ & $	24	$ & $	-1884	$ & $	19	/	18	$ & $	20	$ & $	\quad	$ \\
135	&	 RHN	&	 IO	& $	3	$ & $	-9	$ & $	5	$ & $	0	$ & $	0	$ & $	1	$ & $	6	$ & $	0	$ & $	33	$ & $	-2109	$ & $	1			$ & $	19	$ & $	\quad	$ \\
136	&	 RHN	&	 IO	& $	2	$ & $	-8	$ & $	5	$ & $	0	$ & $	1	$ & $	0	$ & $	3	$ & $	1	$ & $	39	$ & $	-2175	$ & $	65	/	54	$ & $	20	$ & $	\quad	$ \\
137	&	 RHN	&	 IO	& $	2	$ & $	-8	$ & $	5	$ & $	0	$ & $	1	$ & $	1	$ & $	3	$ & $	0	$ & $	42	$ & $	-2460	$ & $	23	/	18	$ & $	20	$ & $	\quad	$ \\
138	&	 RHN	&	 IO	& $	2	$ & $	-8	$ & $	5	$ & $	1	$ & $	0	$ & $	0	$ & $	3	$ & $	0	$ & $	33	$ & $	-2595	$ & $	11	/	10	$ & $	20	$ & $	\quad	$ \\
139	&	 RHN	&	 IO	& $	3	$ & $	-9	$ & $	5	$ & $	1	$ & $	0	$ & $	0	$ & $	6	$ & $	0	$ & $	42	$ & $	-2856	$ & $	27	/	26	$ & $	19	$ & $	\quad	$ \\
140	&	 RHN	&	 IO	& $	2	$ & $	-7	$ & $	5	$ & $	1	$ & $	0	$ & $	1	$ & $	3	$ & $	0	$ & $	24	$ & $	-1524	$ & $	31	/	30	$ & $	20	$ & $	\quad	$ \\
141	&	 RHN	&	 IO	& $	3	$ & $	-9	$ & $	5	$ & $	1	$ & $	0	$ & $	1	$ & $	6	$ & $	1	$ & $	39	$ & $	-2841	$ & $	27	/	26	$ & $	19	$ & $	\quad	$ \\
142	&	 RHN	&	 IO	& $	2	$ & $	-8	$ & $	5	$ & $	1	$ & $	1	$ & $	1	$ & $	3	$ & $	1	$ & $	48	$ & $	-3192	$ & $	13	/	10	$ & $	20	$ & $	\quad	$ \\
143	&	 RHN	&	 IO	& $	2	$ & $	-7	$ & $	5	$ & $	2	$ & $	0	$ & $	0	$ & $	3	$ & $	0	$ & $	33	$ & $	-2199	$ & $	71	/	66	$ & $	20	$ & $	\quad	$ \\
144	&	 RHN	&	 IO	& $	3	$ & $	-8	$ & $	5	$ & $	2	$ & $	0	$ & $	0	$ & $	6	$ & $	0	$ & $	42	$ & $	-2352	$ & $	43	/	42	$ & $	19	$ & $	\quad	$ \\
145	&	 RHN	&	 IO	& $	2	$ & $	-7	$ & $	5	$ & $	2	$ & $	0	$ & $	1	$ & $	3	$ & $	1	$ & $	30	$ & $	-2166	$ & $	71	/	66	$ & $	20	$ & $	\quad	$ \\
146	&	 RHN	&	 IO	& $	2	$ & $	-7	$ & $	5	$ & $	3	$ & $	0	$ & $	0	$ & $	3	$ & $	1	$ & $	39	$ & $	-2913	$ & $	10	/	9	$ & $	20	$ & $	\quad	$ \\
147	&	 RHN	&	 IO	& $	2	$ & $	-7	$ & $	5	$ & $	3	$ & $	0	$ & $	1	$ & $	3	$ & $	0	$ & $	42	$ & $	-3252	$ & $	7	/	6	$ & $	20	$ & $	\quad	$ \\
148	&	 RHN	&	 IO	& $	4	$ & $	-8	$ & $	5	$ & $	3	$ & $	1	$ & $	0	$ & $	6	$ & $	0	$ & $	42	$ & $	-3234	$ & $	91	/	90	$ & $	19	$ & $	\quad	$ \\
149	&	 RHN	&	 IO	& $	3	$ & $	-6	$ & $	5	$ & $	3	$ & $	1	$ & $	1	$ & $	3	$ & $	0	$ & $	24	$ & $	-1740	$ & $	1			$ & $	20	$ & $	\quad	$ \\
150	&	 RHN	&	 IO	& $	4	$ & $	-8	$ & $	5	$ & $	3	$ & $	1	$ & $	1	$ & $	6	$ & $	1	$ & $	39	$ & $	-3201	$ & $	91	/	90	$ & $	19	$ & $	\quad	$ \\
\hline\hline
\end{tabular}
\end{table}

\begin{table}[htb]
\caption{The solution of the anomaly free flavor charge (conti.3)}
\label{tb:fcharge4}
\footnotesize
\centering
\begin{tabular}{ccccccccccccccccc}\hline\hline
No.	& SS & Hierarchy & $k_{Q_1}$ & $k_{H_d}$ & $k_{N_3}$ & $x$ & $y$ & $z$ & $w$ & $p$ & $\Omega$ & $\Xi$ & $k_Y$ &   $\rho$ & $0\nu\beta\beta$ \\ \hline	
151	&	 Neutralino	&	 IO	& $	2	$ & $	-7	$ & $	1	$ & $	0	$ & $	0	$ & $	0	$ & $	3	$ & $	0	$ & $	33	$ & $	-1731	$ & $	61	/	54	$ & $	20	$ & $	\quad	$ \\
152	&	 Neutralino	&	 IO	& $	2	$ & $	-8	$ & $	1	$ & $	0	$ & $	0	$ & $	0	$ & $	3	$ & $	1	$ & $	39	$ & $	-2715	$ & $	65	/	54	$ & $	20	$ & $	\quad	$ \\
153	&	 Neutralino	&	 IO	& $	2	$ & $	-8	$ & $	1	$ & $	0	$ & $	0	$ & $	1	$ & $	3	$ & $	0	$ & $	42	$ & $	-3036	$ & $	23	/	18	$ & $	20	$ & $	\quad	$ \\
154	&	 Neutralino	&	 IO	& $	2	$ & $	-7	$ & $	1	$ & $	0	$ & $	0	$ & $	1	$ & $	3	$ & $	1	$ & $	30	$ & $	-1734	$ & $	61	/	54	$ & $	20	$ & $	\quad	$ \\
155	&	 Neutralino	&	 IO	& $	3	$ & $	-8	$ & $	1	$ & $	0	$ & $	1	$ & $	0	$ & $	3	$ & $	0	$ & $	33	$ & $	-2739	$ & $	61	/	54	$ & $	20	$ & $	\quad	$ \\
156	&	 Neutralino	&	 IO	& $	3	$ & $	-7	$ & $	1	$ & $	0	$ & $	1	$ & $	1	$ & $	3	$ & $	0	$ & $	24	$ & $	-1794	$ & $	19	/	18	$ & $	20	$ & $	\quad	$ \\
157	&	 Neutralino	&	 IO	& $	2	$ & $	-7	$ & $	1	$ & $	1	$ & $	0	$ & $	0	$ & $	3	$ & $	0	$ & $	33	$ & $	-1965	$ & $	11	/	10	$ & $	22	$ & $	\sqrt{}	$ \\
158	&	 Neutralino	&	 IO	& $	2	$ & $	-8	$ & $	1	$ & $	1	$ & $	0	$ & $	0	$ & $	3	$ & $	1	$ & $	39	$ & $	-3039	$ & $	7	/	6	$ & $	22	$ & $	\sqrt{}	$ \\
159	&	 Neutralino	&	 IO	& $	2	$ & $	-7	$ & $	1	$ & $	1	$ & $	0	$ & $	1	$ & $	3	$ & $	1	$ & $	30	$ & $	-1950	$ & $	11	/	10	$ & $	22	$ & $	\sqrt{}	$ \\
160	&	 Neutralino	&	 IO	& $	3	$ & $	-7	$ & $	1	$ & $	1	$ & $	1	$ & $	1	$ & $	3	$ & $	0	$ & $	24	$ & $	-2010	$ & $	31	/	30	$ & $	22	$ & $	\sqrt{}	$ \\
161	&	 Neutralino	&	 IO	& $	3	$ & $	-10	$ & $	2	$ & $	0	$ & $	0	$ & $	0	$ & $	6	$ & $	0	$ & $	42	$ & $	-3252	$ & $	19	/	18	$ & $	22	$ & $	\sqrt{}	$ \\
162	&	 Neutralino	&	 IO	& $	3	$ & $	-9	$ & $	2	$ & $	0	$ & $	0	$ & $	1	$ & $	6	$ & $	0	$ & $	33	$ & $	-2109	$ & $	1			$ & $	22	$ & $	\sqrt{}	$ \\
163	&	 Neutralino	&	 IO	& $	2	$ & $	-7	$ & $	2	$ & $	2	$ & $	0	$ & $	0	$ & $	3	$ & $	1	$ & $	39	$ & $	-2643	$ & $	25	/	22	$ & $	21	$ & $	\sqrt{}	$ \\
164	&	 Neutralino	&	 IO	& $	2	$ & $	-7	$ & $	2	$ & $	2	$ & $	0	$ & $	1	$ & $	3	$ & $	0	$ & $	42	$ & $	-2964	$ & $	79	/	66	$ & $	21	$ & $	\sqrt{}	$ \\
165	&	 Neutralino	&	 IO	& $	3	$ & $	-7	$ & $	2	$ & $	2	$ & $	1	$ & $	0	$ & $	3	$ & $	0	$ & $	33	$ & $	-2703	$ & $	71	/	66	$ & $	21	$ & $	\sqrt{}	$ \\
166	&	 Neutralino	&	 IO	& $	3	$ & $	-6	$ & $	2	$ & $	2	$ & $	1	$ & $	1	$ & $	3	$ & $	0	$ & $	24	$ & $	-1578	$ & $	67	/	66	$ & $	21	$ & $	\sqrt{}	$ \\
167	&	 Neutralino	&	 IO	& $	2	$ & $	-6	$ & $	2	$ & $	3	$ & $	0	$ & $	1	$ & $	3	$ & $	0	$ & $	42	$ & $	-2442	$ & $	7	/	6	$ & $	21	$ & $	\sqrt{}	$ \\
168	&	 Neutralino	&	 IO	& $	3	$ & $	-6	$ & $	2	$ & $	3	$ & $	1	$ & $	0	$ & $	3	$ & $	0	$ & $	33	$ & $	-2235	$ & $	19	/	18	$ & $	21	$ & $	\sqrt{}	$ \\
169	&	 Neutralino	&	 IO	& $	3	$ & $	-6	$ & $	2	$ & $	3	$ & $	1	$ & $	1	$ & $	3	$ & $	1	$ & $	30	$ & $	-2202	$ & $	19	/	18	$ & $	21	$ & $	\sqrt{}	$ \\
170	&	 Neutralino	&	 IO	& $	3	$ & $	-9	$ & $	3	$ & $	0	$ & $	0	$ & $	0	$ & $	6	$ & $	0	$ & $	42	$ & $	-2568	$ & $	19	/	18	$ & $	19	$ & $	\quad	$ \\
171	&	 Neutralino	&	 IO	& $	3	$ & $	-9	$ & $	3	$ & $	0	$ & $	0	$ & $	1	$ & $	6	$ & $	1	$ & $	39	$ & $	-2571	$ & $	19	/	18	$ & $	19	$ & $	\quad	$ \\
172	&	 Neutralino	&	 IO	& $	4	$ & $	-9	$ & $	3	$ & $	0	$ & $	1	$ & $	1	$ & $	6	$ & $	0	$ & $	33	$ & $	-2667	$ & $	1			$ & $	19	$ & $	\quad	$ \\
173	&	 Neutralino	&	 IO	& $	3	$ & $	-9	$ & $	3	$ & $	1	$ & $	0	$ & $	0	$ & $	6	$ & $	0	$ & $	42	$ & $	-2856	$ & $	27	/	26	$ & $	21	$ & $	\sqrt{}	$ \\
174	&	 Neutralino	&	 IO	& $	3	$ & $	-9	$ & $	3	$ & $	1	$ & $	0	$ & $	1	$ & $	6	$ & $	1	$ & $	39	$ & $	-2841	$ & $	27	/	26	$ & $	21	$ & $	\sqrt{}	$ \\
175	&	 Neutralino	&	 IO	& $	2	$ & $	-6	$ & $	3	$ & $	3	$ & $	0	$ & $	1	$ & $	3	$ & $	1	$ & $	48	$ & $	-2904	$ & $	11	/	9	$ & $	18	$ & $	\quad	$ \\
176	&	 Neutralino	&	 IO	& $	3	$ & $	-6	$ & $	3	$ & $	3	$ & $	1	$ & $	0	$ & $	3	$ & $	1	$ & $	39	$ & $	-2679	$ & $	10	/	9	$ & $	18	$ & $	\quad	$ \\
177	&	 Neutralino	&	 IO	& $	3	$ & $	-6	$ & $	3	$ & $	3	$ & $	1	$ & $	1	$ & $	3	$ & $	0	$ & $	42	$ & $	-3000	$ & $	7	/	6	$ & $	18	$ & $	\quad	$ \\
178	&	 Neutralino	&	 IO	& $	3	$ & $	-8	$ & $	4	$ & $	2	$ & $	0	$ & $	0	$ & $	6	$ & $	0	$ & $	42	$ & $	-2352	$ & $	43	/	42	$ & $	20	$ & $	\quad	$ \\
179	&	 Neutralino	&	 IO	& $	4	$ & $	-8	$ & $	5	$ & $	3	$ & $	1	$ & $	0	$ & $	6	$ & $	0	$ & $	42	$ & $	-3234	$ & $	91	/	90	$ & $	19	$ & $	\quad	$ \\
180	&	 Neutralino	&	 IO	& $	4	$ & $	-8	$ & $	5	$ & $	3	$ & $	1	$ & $	1	$ & $	6	$ & $	1	$ & $	39	$ & $	-3201	$ & $	91	/	90	$ & $	19	$ & $	\quad	$ \\
\hline\hline
\end{tabular}
\end{table}
\clearpage 

\bibliography{reference}
\bibliographystyle{JHEP}

\end{document}